\documentclass[aoas,preprint]{imsart}

\usepackage{amsfonts,amstext, mathtools, comment, mathtools} 
\usepackage{natbib}
\usepackage{color}

\usepackage[pdftex]{graphicx}
\usepackage[toc,page]{appendix}
\usepackage{amssymb}
\usepackage{ifthen}
\usepackage{enumitem}
\usepackage{grffile}
 \usepackage{float}

\usepackage{subfig}
\usepackage{subfloat}
\usepackage{multirow}
\usepackage{pifont}

\DeclareMathOperator{\RoI}{RoI}

\graphicspath{{./plots/}}

\newcommand{\cmark}{\ding{51}}%
\newcommand{\xmark}{\ding{55}}

\begin{document}


\begin{frontmatter}
\title{Statistical regionalization for estimation of extreme river discharges}
\runtitle{Statistical regionalization}

\begin{aug}

\author{\fnms{Peiman} \snm{Asadi}\thanksref{m1}\ead[label=e2]{peiman.asadi@epfl.ch}}
\and
\author{\fnms{Sebastian} \snm{Engelke}\thanksref{m1}\ead[label=e1]{sebastian.engelke@epfl.ch}}
\and
\author{\fnms{Anthony C.} \snm{Davison}\thanksref{m1}\ead[label=e3]{anthony.davison@epfl.ch}}

\affiliation{Ecole Polytechnique F\'ed\'erale de Lausanne\thanksmark{m1}}

%

\runauthor{P.~Asadi, S.~Engelke and A.~C.~Davison}
\end{aug}

\begin{abstract}
Regionalization methods have long been used to estimate high return levels of river discharges at ungauged locations on a river network. In these methods, the recorded discharge measurements of a group of similar, gauged, stations is used to estimate high quantiles at the target catchment that has no observations. This group is called the region of influence and its similarity to the ungauged location is measured in terms of physical and meteorological catchment attributes. We develop a statistical method for estimation of high return levels  based on regionalizing the parameters of a generalized extreme value distribution. The region of influence is chosen in an optimal way, ensuring similarity and in-group homogeneity. Our method is applied to discharge data from the Rhine basin in Switzerland, and its performance at ungauged locations is compared to that of classical regionalization methods. For gauged locations we show how our approach improves the estimation uncertainty for long return periods by combining local measurements with those from the region of influence. 
\end{abstract}


\end{frontmatter}

\section{Introduction}

The accurate quantification of high return levels of peak river flows is 
crucial for national agencies, which must design effective flood protection at minimal economic and ecological costs.
For gauging stations with long and stationary discharge records, extreme value
statistics provides reliable tools for model fitting,
identification and assessment of parameter uncertainty.
The two main statistical approaches are the block maximum method, which fits
a generalized extreme value distribution to the yearly maximum discharges, and
the peaks-over-threshold method, which models all exceedances over a high
threshold by a generalized Pareto distribution \citep{kat2002}.
The theory and statistical properties of both methodologies are well-understood and they are used to analyze flood
risk at gauging stations.
 More precisely, if $G_0$ denotes the 
the cumulative distribution function of annual maximum discharge at the target location, and $G_0^{-1}$ its pseudo-inverse, the quantity of interest is the $T$-year return level 
\begin{align}\label{QT}
  Q_0^T = G_0^{-1}(1 - 1/T),
\end{align}
corresponding to a flood that occurs on average once every $T$ years. Typical return
periods $T$ required by regulators are $50$, $100$ or $200$ years, but they may
be much higher at critical sites. 

Flood estimates are often required at locations where no or only few years of observations are available,
or the data quality is inadequate, and then classical tools for high quantile estimation are inapplicable or subject to large estimation uncertainty. 
A common way of estimating the quantiles \eqref{QT} is then to identify gauging stations similar to the target location and transfer information there from
these stations. 
This  is called regionalization. Similarity between locations is measured in geographical distance or based on physical and meteorological catchment attributes. The group of gauging stations similar to the target site is called its region of influence. In general, if the region of influence is well-chosen and contains relevant information on high return levels at target location,
this approach gives good results \citep{grehy1996b}. The identification of 
hydrologically similar stations is as crucial to the final estimation
quality as it is difficult. A wide variety of statistical methods for this have been proposed, including cluster analysis, principal component analysis and canonical correlation analysis \citep{grehy1996a, ilorme2013}.

Once the region of influence is identified, there are two main approaches for transferring the information from the region to the target location. The first approach relies on a hydrological model \citep[e.g.,][]{viviroli2009} whose tunable parameters are calibrated for the gauged catchment and then extrapolated to the possibly ungauged target catchment by the means of nearest neighbors, kriging or regression. Given the estimated parameters, catchment characteristics and meteorological data (observed precipitation, temperature, etc.) the hydrological model then produces the discharge time series
at the location of interest in a deterministic, process-oriented way.  
Hydrological models are appealing as their output includes
the temporal discharge evolution rather than only the marginal
distribution, but they require numerous attributes and 
parameters. Moreover, for estimation of high return levels,
a further extreme-value analysis of the generated time series
is necessary. 

The second approach, called statistical regionalization, directly models
high return levels at the target catchment as a function
of quantities at gauged locations in the region of influence.
The commonest methods in this context are the index flood \citep{dalrymple1960flood} and  quantile regression methods \citep{pandey1999comparative}, but they do not determine the whole distribution function of extremes at the target location and can therefore lead to inconsistencies such as
$Q^{100}_0 < Q^{50}_0$, that is, a 100-year return level that is smaller than the corresponding 50-year return level.

In this paper we propose a new regionalization approach to identify the region of influence
and an estimation approach that solves
the inconsistency issue and reduces the subjectivity involved in other approaches: 
\begin{itemize}
\item
  it regresses the parameters of the generalized extreme value distributions
  at the gauged stations in the region of influence of the target location
  on catchment attributes. The fitted model is applied to
  the target location and provides an estimate of the whole distribution
  of yearly maximum discharges. This directly implies the correct
  ordering of return levels; 
\item
  the similarity measure that determines the distance between catchments
  and therefore the regions of influence is 
  chosen to minimize the estimation error, so no expert
  judgment or heuristics for similarity are required; and 
\item
  the number of required attributes is very small and the computational cost
  compared to hydrological models is considerably reduced.
\end{itemize} 
  
As an application of our method, we estimate high return levels of river discharges
at gauged and ungauged locations on the Rhine basin in Switzerland.
We first use our regionalization approach to improve the at-site 
estimation at gauged target locations by borrowing information from hydrologically similar stations. We compare the results to the classical, local, model that fits an extreme value distribution using data only from the target station.
We then treat each station successively as ungauged and use our method to estimate return levels with return periods up to $200$ years, using the optimal region of influence. The estimations are compared to the quantiles obtained by competing methods such as clustering and canonical correlation analysis in combination with quantile regression. 

In Section~\ref{sec_pre} we recall some notions of univariate extreme value theory and
regionalization. In Section~\ref{sec:meth}, some existing methods are discussed and then our new statistical regionalization approach is introduced. We present 
discharge data at $68$ gauging stations on the Swiss Rhine basin in Section~\ref{Data}
and, in Section~\ref{Results}, we apply our new method to estimate high quantiles at both gauged and ungauged locations.

\section{Extreme values and regionalization}\label{sec_pre}

In this section we give some background on generalized extreme
value distributions, a flexible class of distributions often used to model yearly maxima discharges at gauged locations. We further discuss the notions of region of influence
and hydrological distance, which can be used as a similarity measure for catchments.
Both concepts are crucial in statistical regionalization of river discharges and will 
be used in Section~\ref{sec:meth}.

We consider a river network with $m$ gauged locations
$t_1,\dots ,t_m$, and we denote a generic target location on this network by $t_0$. The target
location $t_0$ might be either gauged or ungauged. The $K$ catchment attributes
that are available at all stations are denoted by $y_{j,k}$ for $j=0,1,\dots, m$ and $k= 1,\dots, K$.

\subsection{Generalized extreme value distributions}\label{local_model}

Generalized extreme value (GEV) distributions have been applied extensively for modeling rare phenomena such as financial crises, heat waves or heavy precipitation. Indeed, the GEV distribution is the only possible limit for the linearly normalized maximum of a sequence of independent, identically distributed random variables \citep{fis1928}. This remains true if the daily observations show temporal dependence, as is often the case in meteorological data \citep{Leadbetter.Lindgren.Rootzen:1983}, so the GEV distribution is the natural choice as a model for yearly maxima of daily river discharges at a gauging station. More precisely, for a gauging station $t_j$, $j=1,\dots, m$, on the river network, we define the local GEV distribution as
\begin{align}
G_{j,L}(x) = 
\begin{cases}
\exp \Big \lbrace  -\Big[1+\xi_j \big (\frac{x-\mu_j}{\sigma_j} \big )\Big]_+^{-1/\xi_j} \Big \rbrace,  & \xi_j \neq 0, \\
\exp \left[-\exp\left\{-\left(\frac{x-\mu_j}{\sigma_j}\right)\right\}\right], & \xi_j = 0,
 \end{cases}
 \label{EqGev}
\end{align}
where $z_{+}=\text{max}(z,0)$ and $\mu_j \in \mathbb{R}$, $\sigma_j > 0 $ and $\xi_j \in \mathbb{R}$ are the location, scale and shape parameters, respectively. The case $\xi_j=0$ in \eqref{EqGev} is obtained  as the limit of the first case when $\xi_j \rightarrow 0$. The three parameters of (\ref{EqGev}) must be estimated from discharge measurements at location $t_j$. For sufficiently long and stationary discharge records there are many well-understood methods for obtaining parameter estimates and confidence intervals; see \citet{kat2002} for more details. Depending on data availability, these methods either use the block maxima of yearly peak discharge values, or all daily discharges that exceed a high threshold, the so-called peaks-over-threshold. Both approaches provide estimates for the parameters $\mu_j,\sigma_j$ and $\xi_j$ and thus of the $(1-1/T)$-quantile, $T > 1$, of the fitted GEV distribution, by replacing the parameters in 
\begin{align}
Q_{j,L}^T = 
\begin{cases}
\mu_j - \frac{\sigma_j}{\xi_j} \big [ 1- \{-\log (1 - 1/T)\}^{-\xi_j}\big],  & \xi_j \neq 0, \\
\mu_j - \sigma_j \log \{ -\log(1 - 1/T)\}, & \xi_j = 0,
 \end{cases}
  \label{EqGevQuant}
\end{align}
by estimates; this quantile is the return level associated with the return period $T$ years.

\subsection{Region of influence}\label{gen_roi}

Fitting the parameters of the GEV distribution in \eqref{EqGev} requires a long
record of discharge values at the location of interest, but ungauged locations 
on the river network, for example at confluence points, cities, or power plants, are also crucial for risk assessment. Standard extreme value methods no longer apply, however, because of the lack of data at ungauged sites.  Even at gauged locations, the record length is often rather short in comparison to the desired return periods, resulting in very high uncertainty.

Regionalization involves the use of discharge information from other gauged locations on the river network that are similar to the target location $t_0$. This group of most similar catchments, denoted by  $ \RoI(t_0)$, is called the region of influence of location $t_0$. Once this group is found, information on high return levels of river discharges is transferred from the region of influence to the ungauged location.

Finding a region of influence which has good properties and which is similar to the target location $t_0$ is an area of research in itself. If no discharge measurements are available at $t_0$ the only information on the corresponding catchment are covariates such as the catchment topology, its size, mean elevation or slope, which can be obtained from digital elevation models. Meteorological quantities such as mean precipitation may also be available. Based on these catchment attributes, a similarity measure is defined and those gauged stations out of $t_1,\dots ,t_m$ that are most similar to $t_0$ according to this measure define $ \RoI(t_0)$. The number of stations $J\in \mathbb N$ in $ \RoI(t_0)$ is a tuning parameter.

Similarity of the catchments of $t_0$ and the stations in $ \RoI(t_0)$ is however not the only criterion for a region of influence. In order to fit a statistical model to high return levels of discharges in $ \RoI(t_0)$, it is necessary that the discharge distributions at stations in this group are sufficiently homogeneous, and many tests for this have been proposed \citep[see, e.g.,][]{hosking1993some}.

There is clearly a payoff between similarity and homogeneity. High similarity between the 
target station and its region of influence ensures a strong relation between
discharges at the ungauged site and the gauged locations, whereas a homogeneous
region of influence is the basis for a sensible statistical model and for information transfer.
Our new regionalization method aims to find an optimal compromise between these two properties.

\subsection{Hydrological distance}

In order to identify the gauged catchments that are most similar to the catchment of the target location of interest, we need  a measure of similarity.
Two types of attributes are typically used: those based on physical characteristics including catchment area, mean altitude or mean slope, and those based on meteorological inputs such as mean precipitation.  In addition, measured discharge statistics might be considered at gauged locations. Even if the target is gauged and discharge statistics are available,
\citet{Castel:2001} argue that the use of physical and meteorological catchment characteristics
is preferable to avoid identifying the region of influence and testing its homogeneity on the
basis of the same information.

Most similarity measures between catchments $t_i$ and $t_j$ are defined as a weighted Euclidean distance
\citep{Burn:1990b,Merz:2005}, 
\begin{align}
D_{i,j} = \Big [w_0 d^{\rm Euc}(t_i,t_j)^2+ \sum_{k=1}^{K}w_k(\tilde y_{i,k}-\tilde y_{j,k})^2 \Big ]^{1/2},
\label{EqDistanceMod}
\end{align}
where $d^{\rm Euc}(t_i,t_j)$ is the Euclidean distance between the centroids of the sub-catchments of $t_i$ and $t_j$, and $w_k \geq 0$ is the relative importance of the $k$th attribute. The catchment attributes $\tilde y_{\cdot ,k}$ are first normalized 
to adjust for their different scales \citep[see][]{Gaal:2008,Merz:2005}.
Classically the Euclidean distance was not used in \eqref{EqDistanceMod}, that is, $w_0=0$, but
\citet{Merz:2005} showed that adding spatial proximity for the identification of the region of influence
improves the estimation error of regionalization \citep[see also][]{Gaal:2009}.

\section{Methodology}\label{sec:meth}

\subsection{Methods }\label{met_lit}

The problem of estimating the high quantiles $Q_0^T$ in \eqref{QT} at the target location $t_0$
attracts ongoing attention in the hydrological literature.
Here we describe only the most commonly-used methods that will later be compared to
our new approach presented in Section~\ref{our_method}.

\subsubsection{Regional model}
Unlike the local model described in \S\ref{local_model}, regionalization uses 
additional information from similar catchments in order to decrease the estimation error of the
model. Consider a region $R = \{t_1,\dots, t_J\}$ of $J\in\mathbb N$ gauged locations. In order
to pool the information in $R$ and to use common parameters for high quantiles, a joint 
regional model is required. The most popular regional model  is quantile regression, which assumes that for a fixed return period $T$, the $(1-1/T)$-quantiles of discharge distributions in region $R$ follow a log-linear model \citep{Benson:1962}
\begin{align}
\label{Eq_RegQuant}
\log Q_{j}^T  = \alpha_0 + \sum_{k=1}^K \alpha_k \log y_{j,k} + \epsilon_j, \quad j=1,\ldots,J,
\end{align}
with covariates given by the catchment attributes $y_{j,k}$, and independent, normally distributed regression error terms $\epsilon_j$. The coefficients $(\alpha_0,\alpha_1,\ldots,\alpha_K) $ can be estimated by ordinary least squares. For a target location $t_0$, either gauged or ungauged,
that is similar to the region $R$, the $T$-year return level can then
be computed as
\begin{align*} \hat Q_0^T= \exp\left( \hat \alpha_0 + \sum_{k=1}^K \hat\alpha_k \log y_{0,k}  \right),
\end{align*}
where $(\hat \alpha_0,\dots, \hat \alpha_K)$ are the fitted model parameters.

\subsubsection{Region of influence}\label{lit_roi}
The regional model above can be used to jointly model high quantiles of locations in $R$, but extrapolation to location $t_0$ requires that its region of influence  $ \RoI(t_0)$ contains only locations similar to the target site. Several techniques have been proposed to identify $ \RoI(t_0)$.

Cluster analysis is a classical fixed-region approach that partitions the geographical study area into distinct regions. A target station $t_0$ belongs to exactly one of these regions, which then defines $ \RoI(t_0)$. In order to find the fixed clusters, the distance matrix
$\{D_{i,j} : i,j=1,\dots , m\}$ between all gauged locations is constructed based on the hydrological 
distance defined in \eqref{EqDistanceMod}. In applications, the weights are usually taken to be equal for all attributes or determined by expert knowledge. Then the ungauged location will be assigned to one of the regions.
Several clustering techniques have been applied to construct the homogeneous regions \citep[see for instance][]{rao2006regionalization, chiang2002hydrologic}. The most common technique is the Ward method, which partitions the entire region into clusters of similar sizes and is considered to be well-suited for regionalization \citep[pp. 58--59]{Hosking:1997}. The Ward method is an agglomerative hierarchical algorithm which starts with $m$ clusters, that is, each location is a single cluster, and then successively merges the two clusters resulting in the smallest increase in total within-cluster variance. This procedure is continued until the desired number $C\in \mathbb N$ of clusters is attained, where $C$ is a tuning parameter. For an application of the Ward method in regionalization see \citet{acreman1986}.

While cluster analysis partitions the area into distinct and fixed regions, canonical correlation analysis
is used to find a tailor-made region of influence $ \RoI(t_0)$ for each target location $t_0$ \citep{hardoon2004}. The idea is to determine a joint normal model for linear combinations of discharge and catchment characteristics with maximal correlations. The assumption of multivariate normality
ensures that the distribution of the discharge characteristics conditioned on the catchment characteristics is also normally distributed with, by construction, diagonal covariance matrix $\Sigma$.
Given the catchment characteristics at the target location $t_0$, the distance between the discharge
distribution at $t_0$ and any observed discharge at gauged locations can then be measured by the Mahalanobis distance induced by $\Sigma$. All stations within a certain radius $r>0$ with respect to this distance form $ \RoI(t_0)$; $r$ is a tuning parameter. \citet{grehy1996b} compared various regionalization approaches and concluded that canonical correlation analysis is generally preferable to other approaches.   See \citet{ouarda2001regional, ouarda2008intercomparison} for applications of canonical correlation analysis.

\subsection{Regionalization of extreme value distributions}\label{our_method}

\subsubsection{Regional GEV distribution}
Similar to classical quantile regression in \eqref{Eq_RegQuant}, for a fixed region $R = \{t_1,\dots, t_J\}$, our method uses common parameters to increase the accuracy of rare event estimation, but instead of modeling a particular quantile, we regionalize the entire distribution function.
This ensures the correct ordering of estimated quantiles for different return levels and avoids inconsistencies encountered with quantile regression.
More precisely, we assume a log-linear relationship between the location and scale parameters of the GEV distributions at different locations in region $R$ and their catchment characteristics.
The shape parameters are taken to be constant over $R$, which is natural if the region is sufficiently homogeneous. See \citet{Cooley:2007} and \citet{Sang:2009} for discussions of the difficulty of finding reliable spatial patterns for the shape parameters of GEV distributions fitted to environmental data. For each station $t_j~(j=1,\ldots,J)$, we define the regionalized GEV distribution with location, scale and shape parameters, respectively, as
\begin{equation}
\begin{aligned}
G_{j,R}(x) &= \exp\left\{- \left(1 + \xi_j \frac{x - \mu_j}{\sigma_j}\right)_+^{-1/\xi_j} \right\},\\
\log\mu_{j} &= \alpha_0 + \sum_{k=1}^{K}\alpha_k \log y_{j,k},\\
\log\sigma_{j} &= \beta_0 + \sum_{k=1}^{K}\beta_k \log y_{j,k},\\
\xi_j&=\xi, 
        \end{aligned}
\label{EqGEVCov}
\end{equation} 
where $(y_{j,1},\ldots , y_{j,K})$ are $K$ covariates at gauging station $j$, and $(\alpha_0,\alpha_1,\ldots,\alpha_K)$ and $(\beta_0, \beta_1,\ldots,\beta_K)$ are the coefficients associated to the covariates.
This marginal model can be fitted by maximizing an independence likelihood \citep{Chandler.Bate:2007}.
An alternative that correctly takes the spatial dependence between the locations into account would be
to fit a max-stable dependence model and optimize the joint likelihood \citep{eng2014b}. Since we are 
only interested in marginal estimates, the much simpler independence likelihood is preferable.

For a target location $t_0$, similar to region $R$, with catchment attributes $(y_{0,1}, \dots, y_{0,K})$,
the estimated $T$-year return level can then be computed as $G^{-1}_{0,R}(1-1/T)$ obtained from model \eqref{EqGEVCov} for the fitted parameters $(\hat \alpha_0,\ldots,\hat\alpha_K)$ and $(\hat\beta_0,\ldots,\hat\beta_K)$ and~$\hat\xi$.

\subsubsection{Optimal region of influence}\label{opt_roi}

Given a target location $t_0$ at which flood estimation is desired,
different methods to determine its region of influence $ \RoI(t_0)$ were discussed in
Section~\ref{lit_roi}. As discussed in Section~\ref{gen_roi}, the stations in $ \RoI(t_0)$ should be both homogeneous and similar to location $t_0$.

To ensure the latter, we can use the hydrological distance in \eqref{EqDistanceMod} with weights
$w_0,w_1,\dots, w_K\geq 0$ satisfying $\sum_{k=0}^K w_k =1$. For a fixed set of weights and
a fixed group size $J\in \mathbb N$, a region $R(\{w_k\}, J)$ can then be defined as the
$J$ gauging stations out of all $\{t_1,\dots, t_m\}$ that are closest to $t_0$ measured in hydrological distance; these are the $J$ nearest neighbors. Once this region is found,
we can fit the model \eqref{EqGEVCov} described above and obtain estimates
of arbitrary quantiles at location $t_0$. Such estimates will, however, strongly depend
on the choice of weights and the group size. The region $R(\{w_k\}, J)$ might be similar  to the target station,
with respect to the chosen distance, but need not also be homogeneous. 

Below we describe how to choose the optimal set of weights
and the optimal group size. Roughly speaking, we range over all regions $R(\{w_k\}, J)$ and minimize the training error,
that is, the estimation error made at stations in $R(\{w_k\}, J)$, whose discharge characteristics
are known from observations. The discharge distributions in the optimal region $R(\{w_k^*\}, J^*)$ are homogeneous in the sense that their high quantiles are explained by the fitted regional model \eqref{EqGEVCov} better than for any other region $R(\{w_k\}, J)$. In order to avoid overfitting and to enforce a minimal
contribution to the similarity measure from all attributes, we impose lower bounds $\epsilon>0$ on the weights and $N \in\mathbb N$ on the group size. The full procedure for finding the optimal region of influence of $t_0$ is the following:
\begin{itemize}
\item
  For all $w_0,\dots, w_K\geq \epsilon$, $\sum_{k=0}^K w_k = 1$, and $J\geq N$,
  \begin{itemize}
    \item
      find the region $R(\{w_k\}, J)$ as the $J$ nearest neighbors of $t_0$;
    \item
      fit the regional model \eqref{EqGEVCov} to $R(\{w_k\}, J)$ and compute the fitted model quantiles
      $\widehat Q^T_j  = \widehat Q^T_j(\{w_k\}, J)$ for all $t_j \in R(\{w_k\}, J)$.
  \end{itemize}
\item
  Compute the optimal weights and group size minimizing the training error, that is,
  \begin{align}
    (w_0^*,\dots, w_K^*, J^*) = \underset{\omega_0,\ldots,\omega_K, J}{\text{argmin}} \,\, \frac{1}{\sum_{j=1}^J N_j}  \sum_{j=1}^J \sum_{T=1}^{N_j}\Bigg( \frac{\widehat Q^{T}_j - X_{T:N_j}(t_j)}{X_{T:N_j}(t_j)}\Bigg)^2,
    \label{Eq_Opt}
  \end{align}
  where $N_j$ is the number of yearly maxima available at gauging station $t_j$ and $X_{T:N_j}(t_j)$ is the corresponding $T$th order statistic. The latter can be seen as the empirical $(1-1/T)$ quantile.
  \item
    The region $ \RoI(t_0) = R(\{w_k^*\}, J^*)$ is the optimal region of influence of location~$t_0$.
  \item
    Use the regional model \eqref{EqGEVCov} fitted to $ \RoI(t_0)$ to estimate the $T$-year return level
    $\widehat Q_0^T$ at target location $t_0$.    
\end{itemize}

Like canonical correlation analysis, our method also defines an individual region 
of influence for each target location. In contrast to methods such as cluster analysis,
we optimize over all possible hydrological distances. The importance of different catchment
attributes is thus chosen automatically, which avoids the use of heuristics or expert judgment. 

Homogeneity inside the region of influence in our approach is not defined
in terms of distribution functions, as is the case for the test of \citet{hosking1993some}.
Instead, the optimization \eqref{Eq_Opt} chooses the most homogeneous region in the sense that
high quantiles can be well estimated by using information in this region. Therefore, homogeneity
is tested using gauged discharge characteristics, whereas the similarity for identifying the
region of influence is based on physical and meteorological attributes only. This agrees
with the reasoning of \citet{Castel:2001}.

\section{Data}\label{Data}
\begin{figure}[t]
\centering
  \includegraphics[trim = 10mm 10mm 5mm 5mm, clip,width= .98\textwidth]{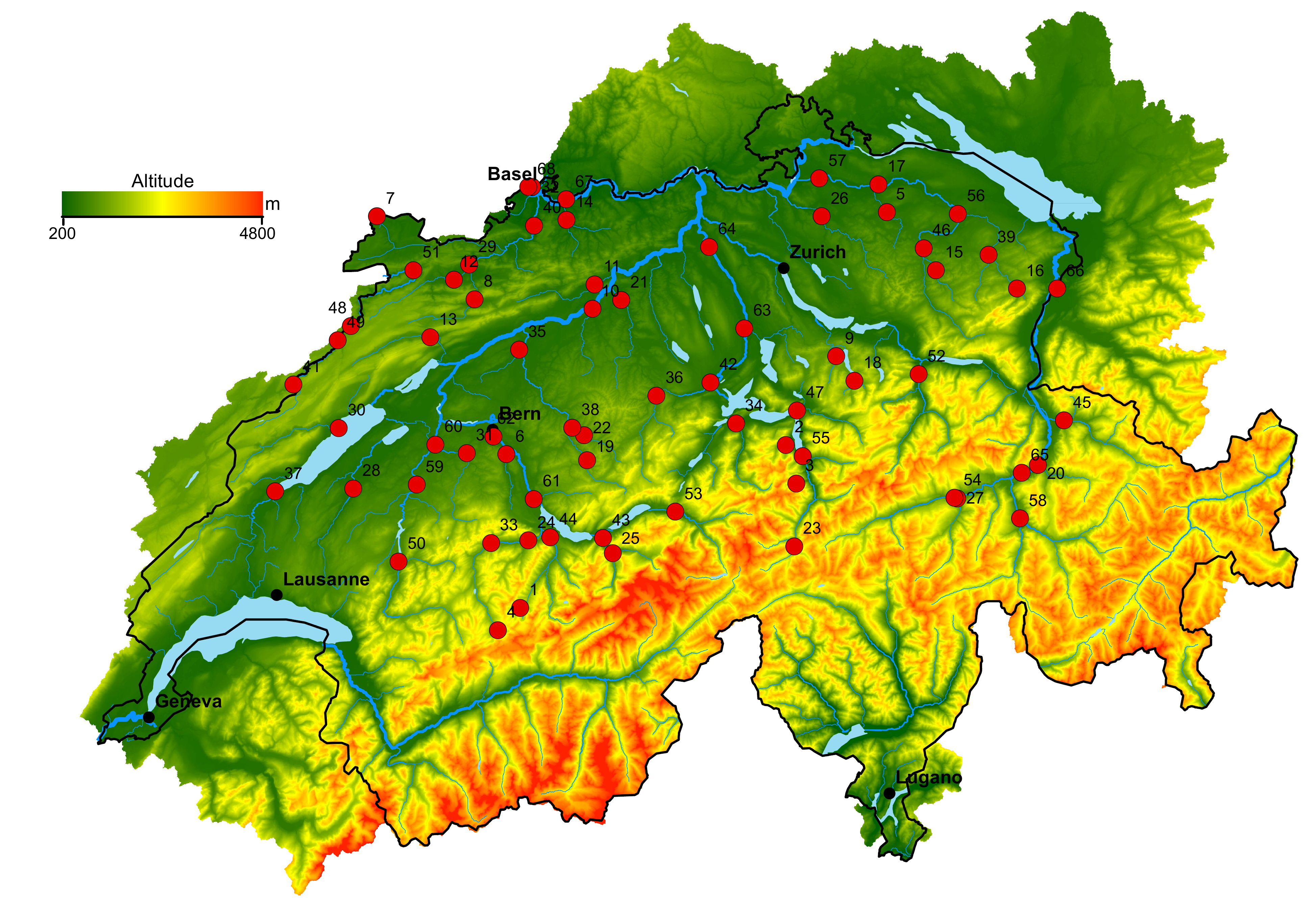}
  \caption{Topographic map of the Rhine basin in Switzerland, showing sites of 68 gauging stations (red dots) along the Rhine, the Aare and their tributaries.}
  \label{Fig_MapSwiss}
\end{figure}

In order to illustrate the new regionalization method developed in the previous section,
we use annual maximum discharge measured at $m=68$ gauging stations $t_1,\dots, t_m$ on $44$ rivers in the Rhine basin in Switzerland, provided by the Swiss Federal Office for the Environment. The time series of annual maxima at station $t_j~(j=1,\dots, m)$, is denoted
by $X_i(t_j)~(i=1,\dots, N_j)$, where $N_j$ denotes the number of available years at
station $t_j$. The lengths of the time series are between 30 and 120 years with an average of 60 years; see the top left panel of Figure~\ref{Fig_DataLength}.
The average discharge of annual maxima at these stations ranges from $18 \text{m}^3/\text{s}$ in the Alps to $2975 \text{m}^3/\text{s}$ at the most downstream station on the Rhine. The major part of the run-off in the basin arises from the Alps.  Figure~\ref{Fig_MapSwiss} shows the boundary of the basin with its topography and the $68$ gauged stations.

In river flow data, seasonality, temporal trend (possibly due to climate change) and abrupt changes due to man-made structures like dams or power plants are the main sources of non-stationarity. As we use only the annual maxima, we can ignore seasonality. Many studies have focused on detecting abrupt changes or trends in extreme river flows. \citet{kun2005} and \citet{Birsan:2005} analyze rivers in central Europe, but neither finds significant trend. In particular, the latter investigates $48$ gauging stations in Switzerland and detects no significant trend in extremes of river flows. To confirm their results we introduce a temporal linear trend 
\citep[see, e.g.,][Ch.~6]{col2001} for the location and scale parameters of the GEV distribution fitted to the annual maxima of the times series. In almost all stations no significant trend is observed, so we treat the data as temporally stationary.

In addition to observed time series of annual maximum discharge, we use a digital elevation model for our study area\footnote{Downloaded from \texttt{http://gdem.ersdac.jspacesystems.or.jp}, made publicly available by the Ministry of Economy, Trade, and Industry of Japan and the United States National Aeronautics and Space Administration; see \citet{Tachikawa:2011}.}. 
The package ARCGIS\footnote{Software for analyzing geographical information and digital elevation models.} is used to obtain the catchment attributes listed in the first part of Table~\ref{Tab_Covariates}. 
The CORINE data base\footnote{Downloaded from \texttt{http://www.eea.europa.eu}} \citep{bossard2000} provides the type of land-cover of each catchment, that is, the portion of forest, rock and agriculture.
We also compute spatio-temporal averages of daily precipitation and averages of annual maximum daily precipitation over each catchment by using its boundaries and historical precipitation data\footnote{Available on \texttt{http://www.euro4m.eu/datasets.html}} on a  5km$\times$5km  grid \citep{isotta2014}. Exploratory analysis shows strong correlation between certain attributes,
and Table~\ref{Tab_Covariates} shows which of them carry relevant information on extremes discharges and 
will be used as covariates in our model.

\begin{table}[t]
\begin{center}
  \begin{tabular}{ lcll}
    \hline
     
    Covariate & Unit & Average &Used \\ \hline
    Physiography & & & \\
    \ \ \ \  Coordinates of the catchment centroid & & &\cmark\\
   \ \ \ \ Size & $km^2$ & 1764 &\cmark\\
   \ \ \ \ Average altitude &$m$&1285&\cmark\\
   \ \ \ \ Average slope &Degree&19&\xmark\\
   \ \ \ \ Density &$km/km^2$&0.7&\xmark\\
   Precipitation & &  \\
   \ \ \ \ Average of daily precipitation &$mm$&4.19&\cmark\\
   \ \ \ \ Average of annual maximum daily precipitation&$mm$&58&\cmark\\
   Land cover type& && \\
   \ \ \ \ Agriculture&$\%$ &26&\xmark\\
  \ \ \ \ Forest& $\%$ &46 &\xmark\\
  \ \ \ \ Rock& $\%$ &16 &\xmark\\
    \hline
      \end{tabular}
  \caption{Summary of the catchment attributes that are available for the study area.}
          \label{Tab_Covariates}
\end{center}
\end{table}

\section{Results}\label{Results}

\subsection{At-site regionalization}

\begin{figure}[!t]
\captionsetup[subfigure]{labelformat=empty}
\subfloat[]{%
  \includegraphics[trim = 0mm 18.5mm 3mm 6mm, clip,width= .45\textwidth]{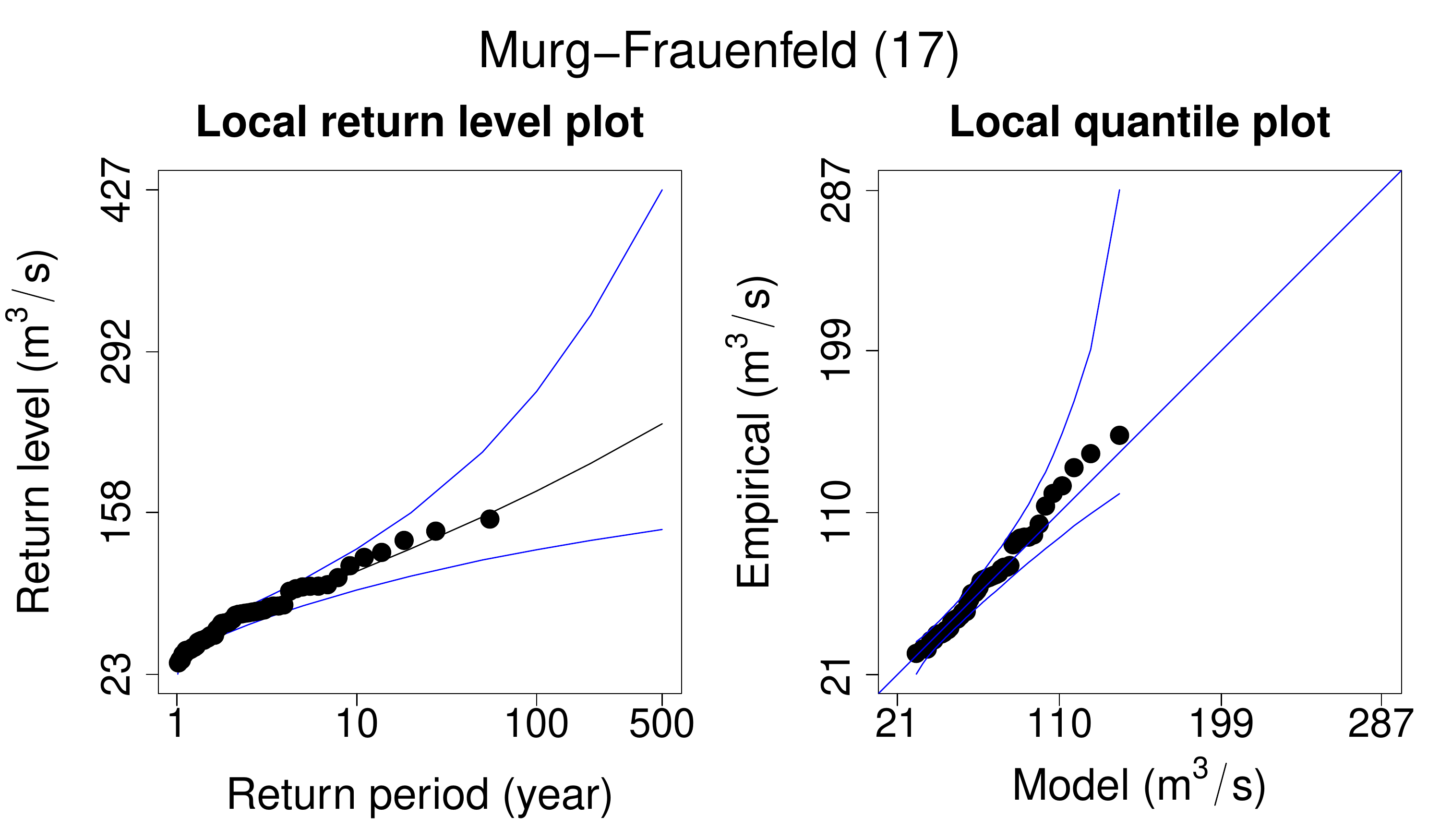}%
  \captionsetup{labelformat=empty}
}
\hspace{1em}
\subfloat[]{%
  \includegraphics[trim =0mm 18.5mm 3mm 6mm, clip,width= .45\textwidth]{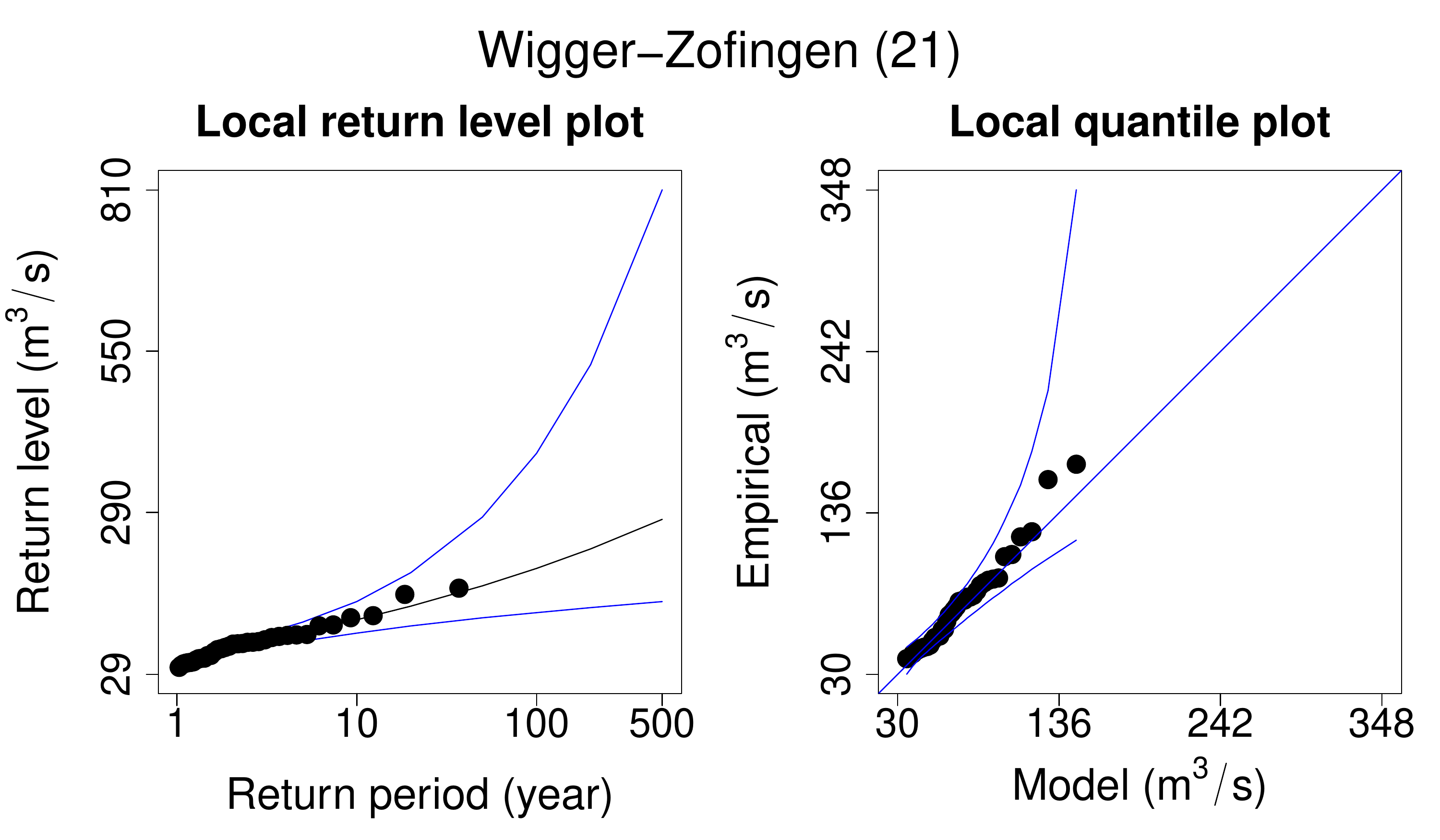}%
  \captionsetup{labelformat=empty}
}
\vspace{-1.5em}
\subfloat[]{%
  \includegraphics[trim = 0mm 5mm 3mm 18mm, clip,width= .45\textwidth]{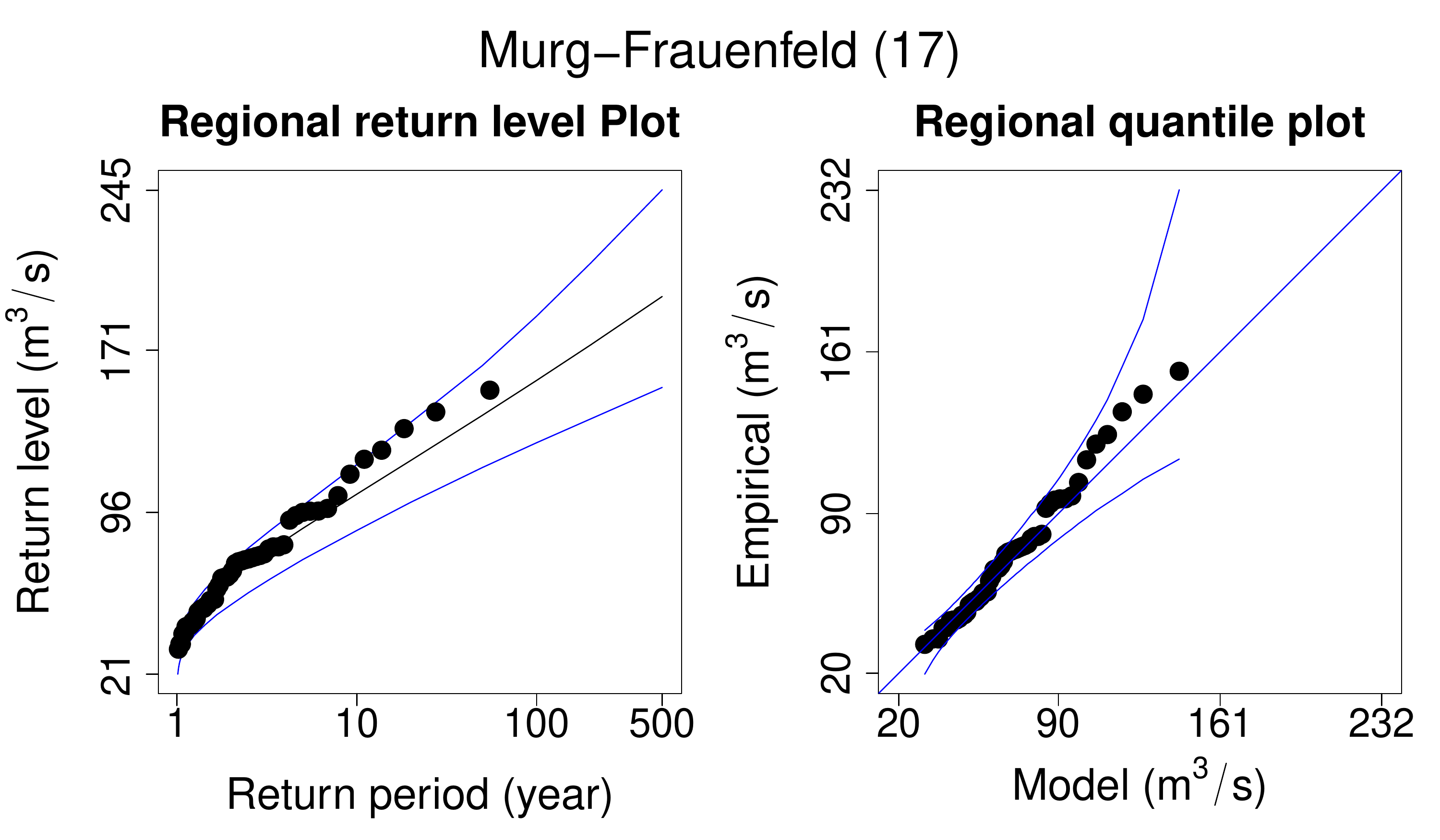}%
  \captionsetup{labelformat=empty}
}
\hspace{1em}
\subfloat[]{%
  \includegraphics[trim = 0mm 5mm 3mm 18mm, clip,width= .45\textwidth]{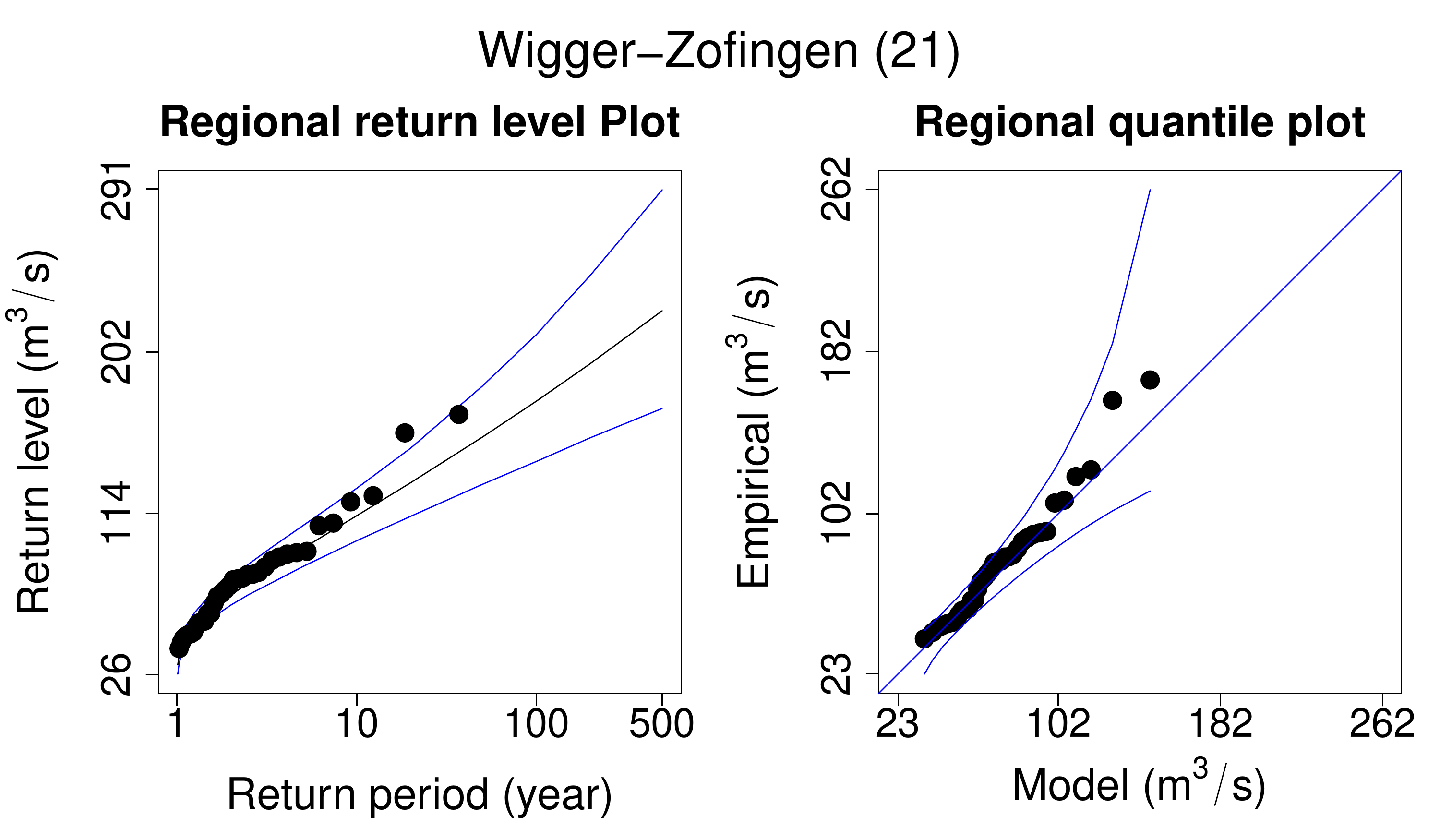}%
  \captionsetup{labelformat=empty}
}

\vspace{-1em}

\subfloat[]{%
  \includegraphics[trim = 0mm 18.5mm 3mm 6mm, clip,width= .45\textwidth]{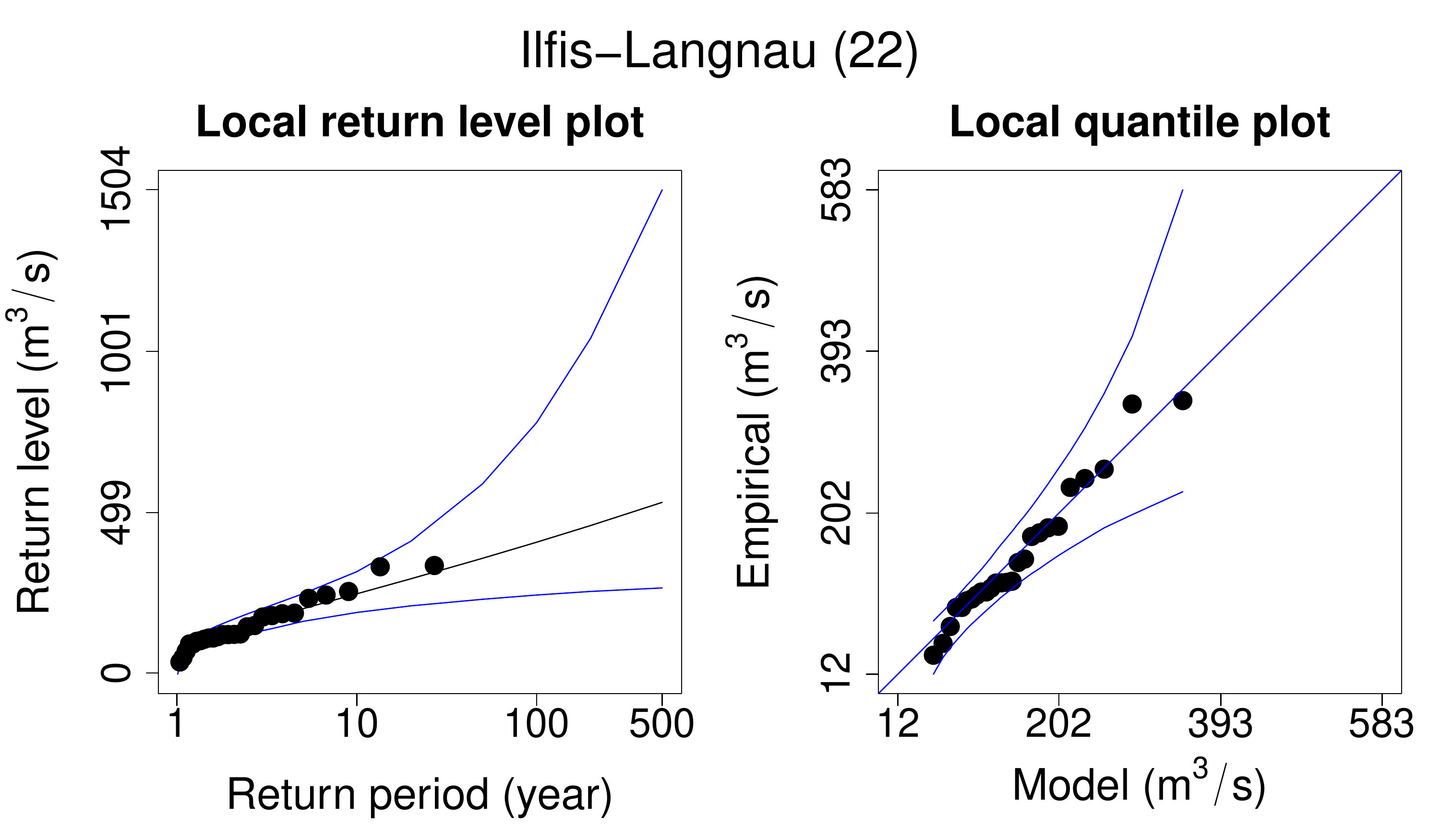}%
  \captionsetup{labelformat=empty}
}
\hspace{1em}
\subfloat[]{%
  \includegraphics[trim =0mm 18.5mm 3mm 6mm, clip,width= .45\textwidth]{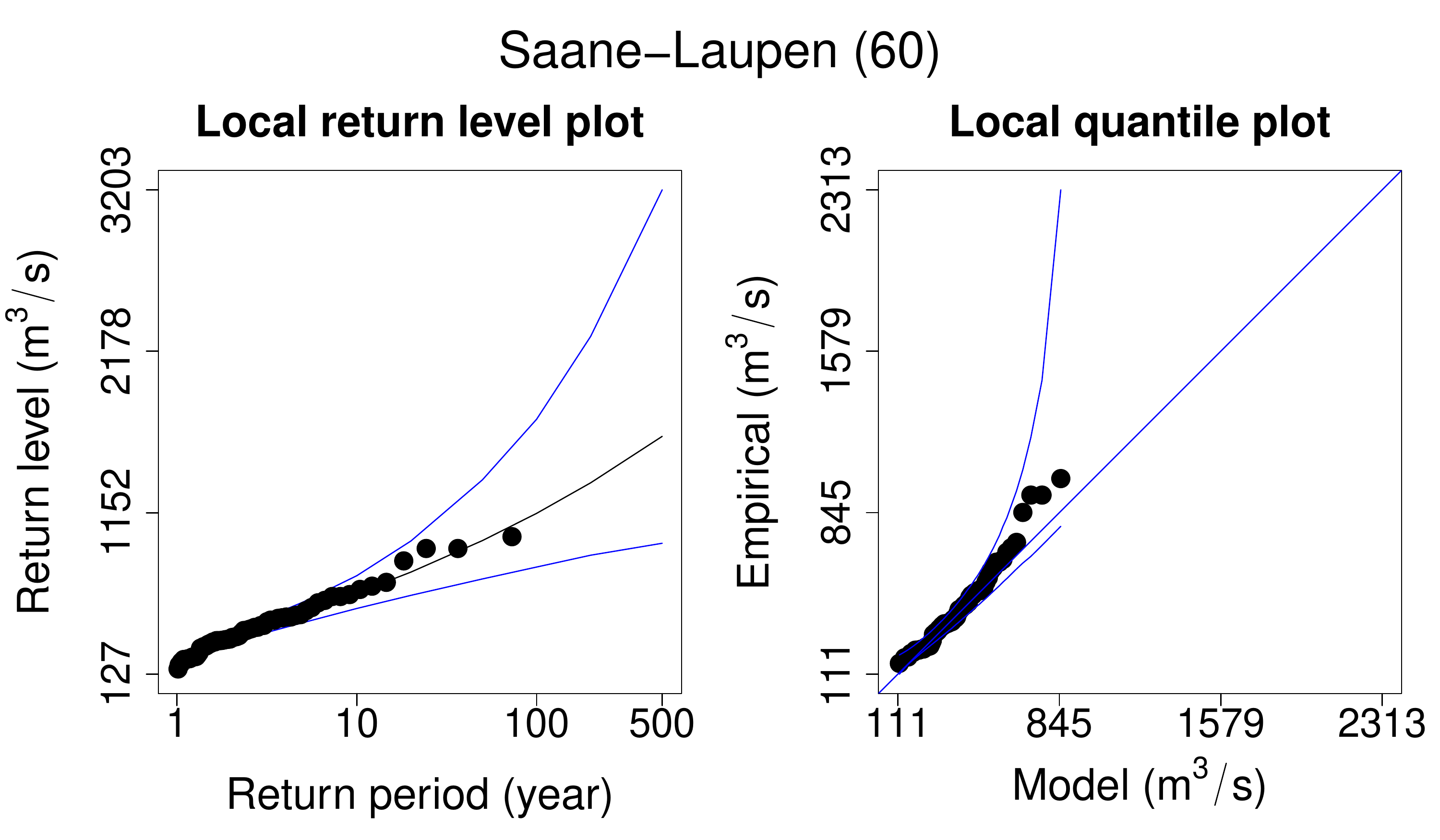}%
  \captionsetup{labelformat=empty}
}
\vspace{-1.5em}
\subfloat[]{%
  \includegraphics[trim = 0mm 5mm 3mm 18mm, clip,width= .45\textwidth]{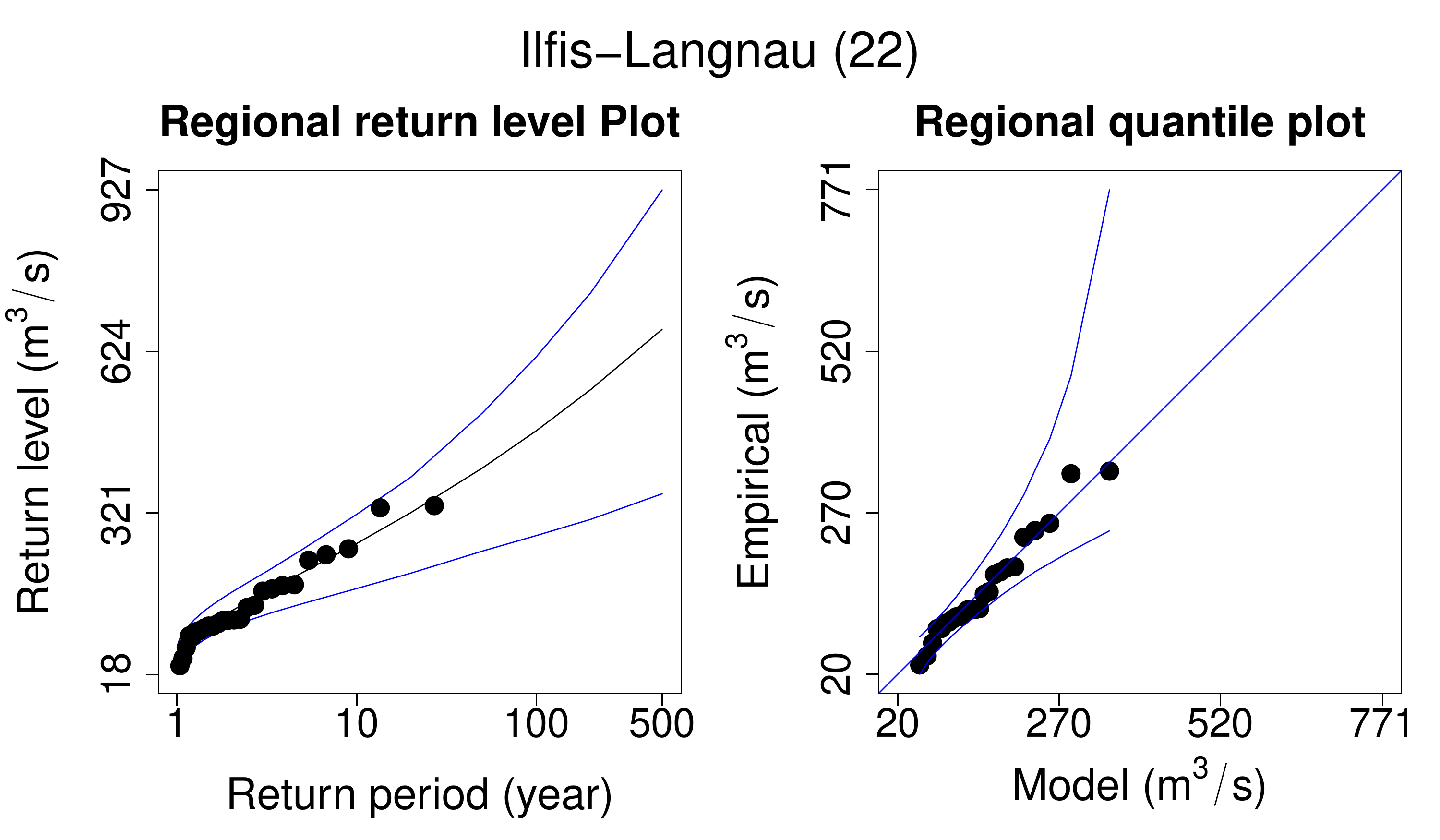}%
  \captionsetup{labelformat=empty}
}
\hspace{1em}
\subfloat[]{%
  \includegraphics[trim = 0mm 5mm 3mm 18mm, clip,width= .45\textwidth]{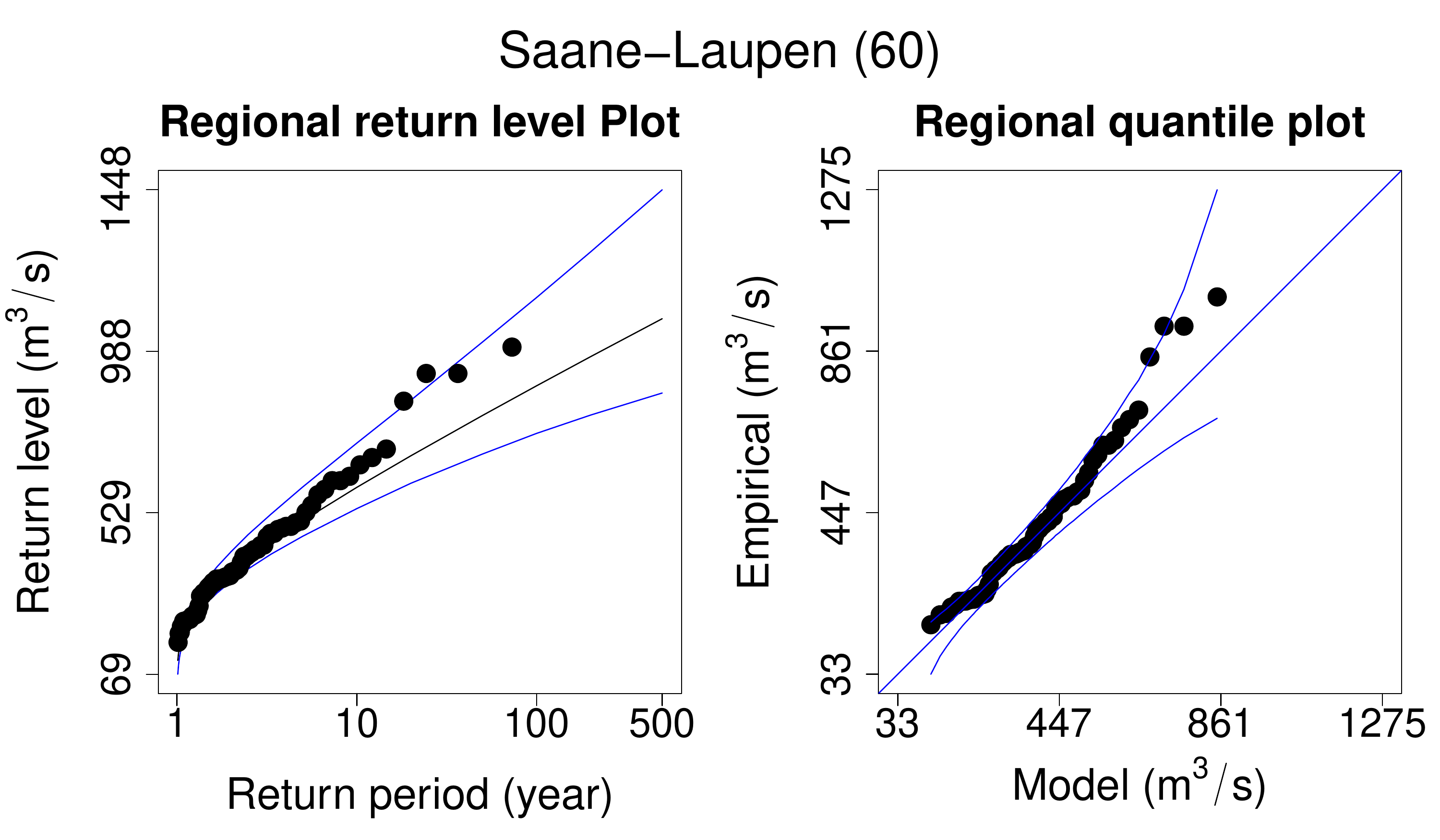}%
  \captionsetup{labelformat=empty}
}
\caption{QQ-plots and return level plots for four stations.  The numbers in the titles show the number of the station as given on the map in Figure~\ref{Fig_MapSwiss}. In each group of four panels, the top and bottom plots show the outputs of local and regional models, the left panels show the return level plots, with blue lines giving pointwise $95\%$ confidence intervals \citep[\S4.2.4]{Davison.Hinkley:1997}, and the right panels compare the data with the fitted models.}
\label{Fig_QQplots}
\end{figure}

In this section we consider the estimation of high quantiles
at a target location $t_0$ with at-site discharge measurements. Standard methods can be used to fit a local GEV distribution to the annual
maxima at this station and to extrapolate into high quantile regions; see Section~\ref{local_model}. Confidence bands for the estimates are typically computed by 
parametric bootstrap \citep{Davison.Hinkley:1997}. The local GEV fits for four stations can be found in the
upper panels of Figure~\ref{Fig_QQplots}. We notice that the local fit accurately 
reproduces the observations, but for higher return periods outside the range of the
data the uncertainty becomes huge and might be unrealistic.

In many applications the number of available years $N_0$ at the target gauged
location is small compared to the desired return periods, but at-site regionalization can be used to decrease the estimation uncertainty.
Instead of fitting a local GEV model to discharge data only
from $t_0$, we can use our methodology in Section~\ref{our_method} to determine the region of influence $ \RoI(t_0)$ and fit
a regional model to the augmented region $\{t_0\}\cup  \RoI(t_0)$.
Our method can be directly applied with a slight modification of the error function
in the minimization~\eqref{Eq_Opt}. Since we have actual at-site discharge measurement
at the target location $t_0$,  the error function in \eqref{Eq_Opt} is replaced by
\begin{align}
  \frac{1}{\sum_{j=0}^J N_j}  \left\{ \tau \sum_{T=1}^{N_0}\Bigg( \frac{\widehat Q^{T}_0 - X_{T:N_0}(t_0)}{X_{T:N_0}(t_0)}\Bigg)^2 +  \sum_{j=1}^J \sum_{T=1}^{N_j}\Bigg( \frac{\widehat Q^{T}_j - X_{T:N_j}(t_j)}{X_{T:N_j}(t_j)}\Bigg)^2 \right\},
\end{align}
where the parameter $\tau\geq 1$ increases the importance of the error 
at this station compared to those in $ \RoI(t_0)$. A large value of $\tau$
results in higher bias and smaller variance, and conversely for small $\tau$.
For our data the value $\tau=2$ gives a good bias-variance trade-off.
The tuning parameters described in Section~\ref{opt_roi} are put to
$\epsilon = 0.05$ as the minimal weight for an attribute and $N = 7$ as the minimal group size.
An example of the optimal region of influence of a gauging station is shown in Figure~\ref{Fig_ROI}
in the Appendix.

\begin{figure}[!t]
\captionsetup[subfigure]{labelformat=empty}
\subfloat[]{%
   \includegraphics[trim = 0mm 20mm 5mm 15mm, clip,scale=.4]{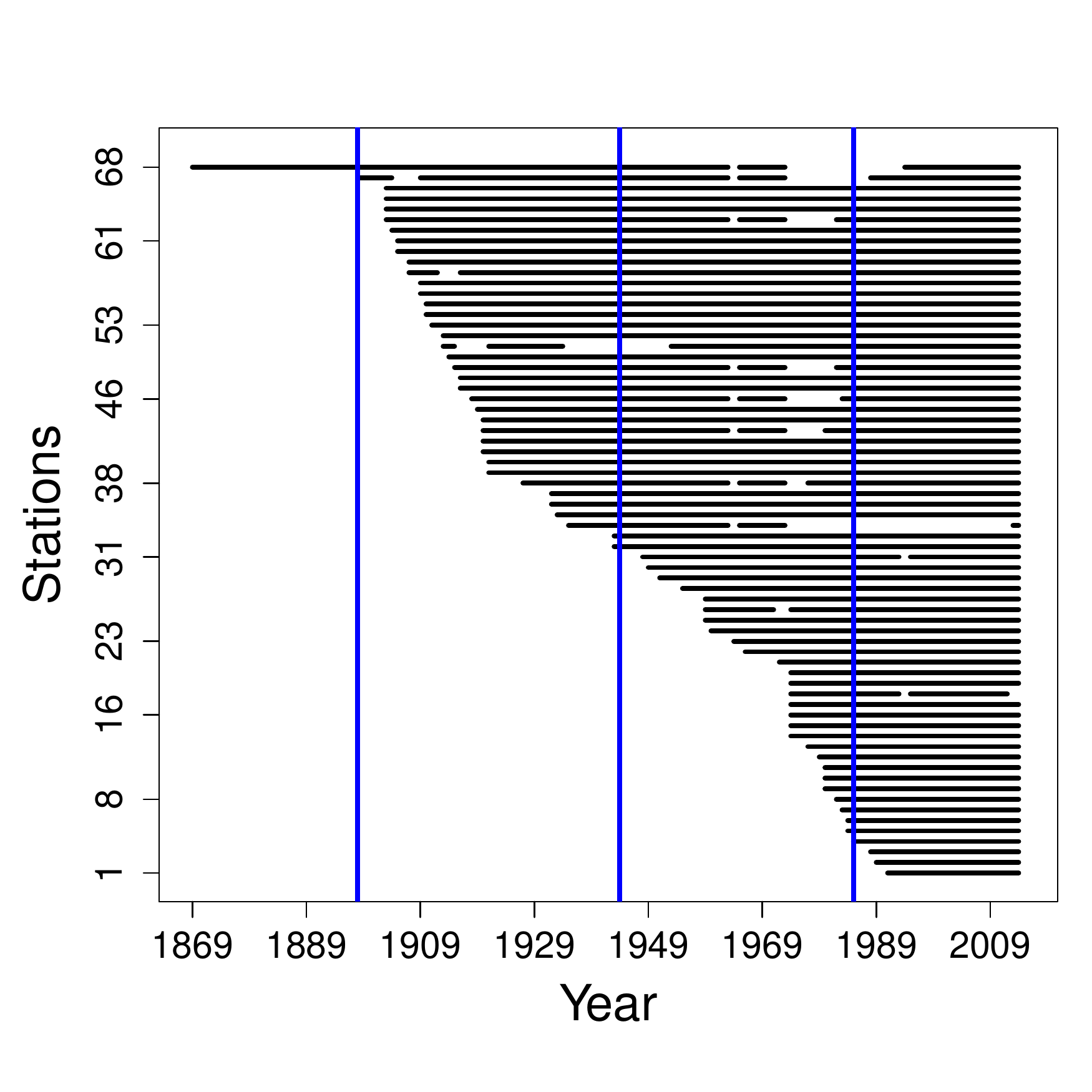}%
  \captionsetup{labelformat=empty}
}
\hspace{0em}
\subfloat[]{%
   \includegraphics[trim = 10mm 20mm 5mm 15mm, clip,scale=.4]{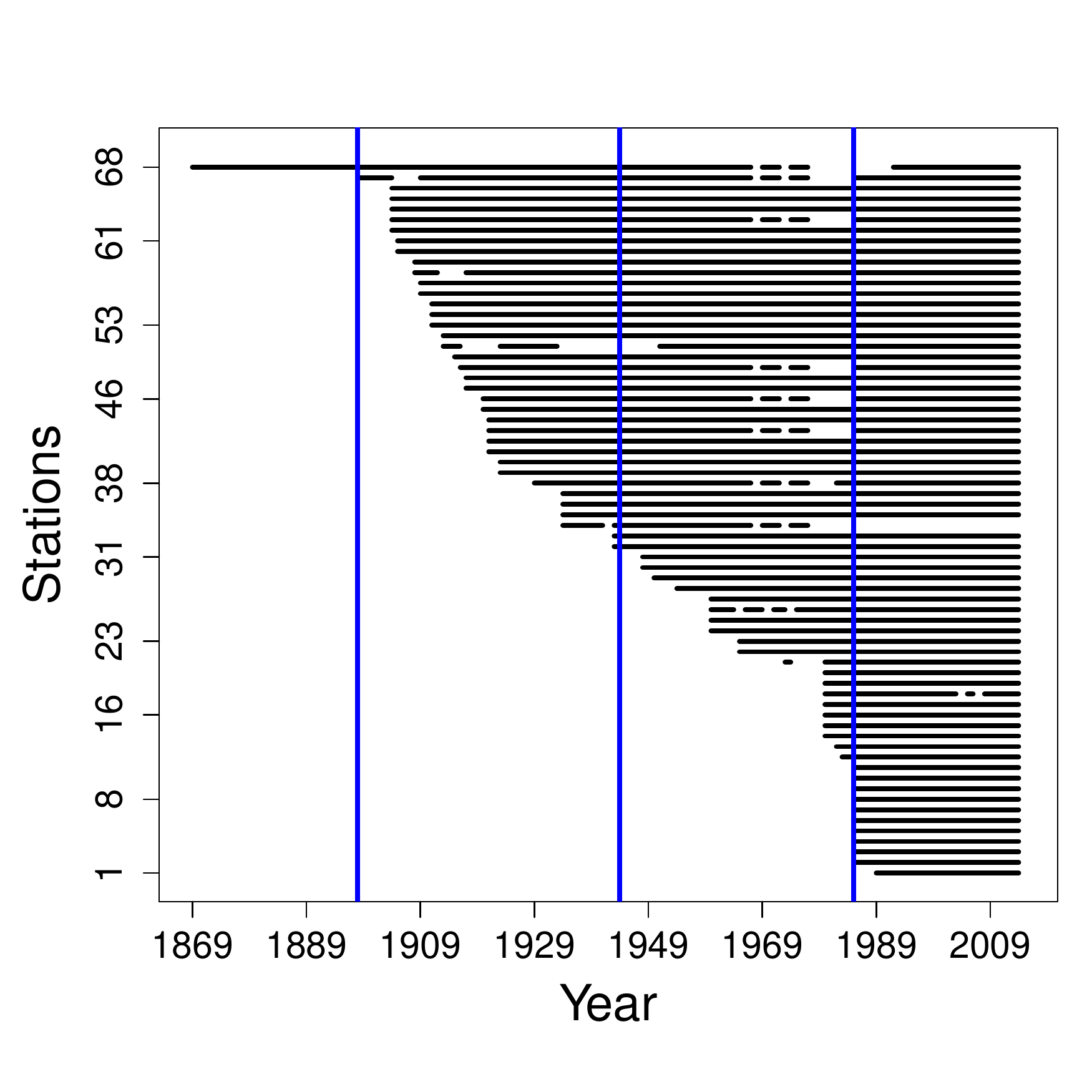}%
  \captionsetup{labelformat=empty}
}

\vspace{-2em}

\subfloat[]{%
   \includegraphics[trim = 0mm 10mm 5mm 15mm, clip,scale=.4]{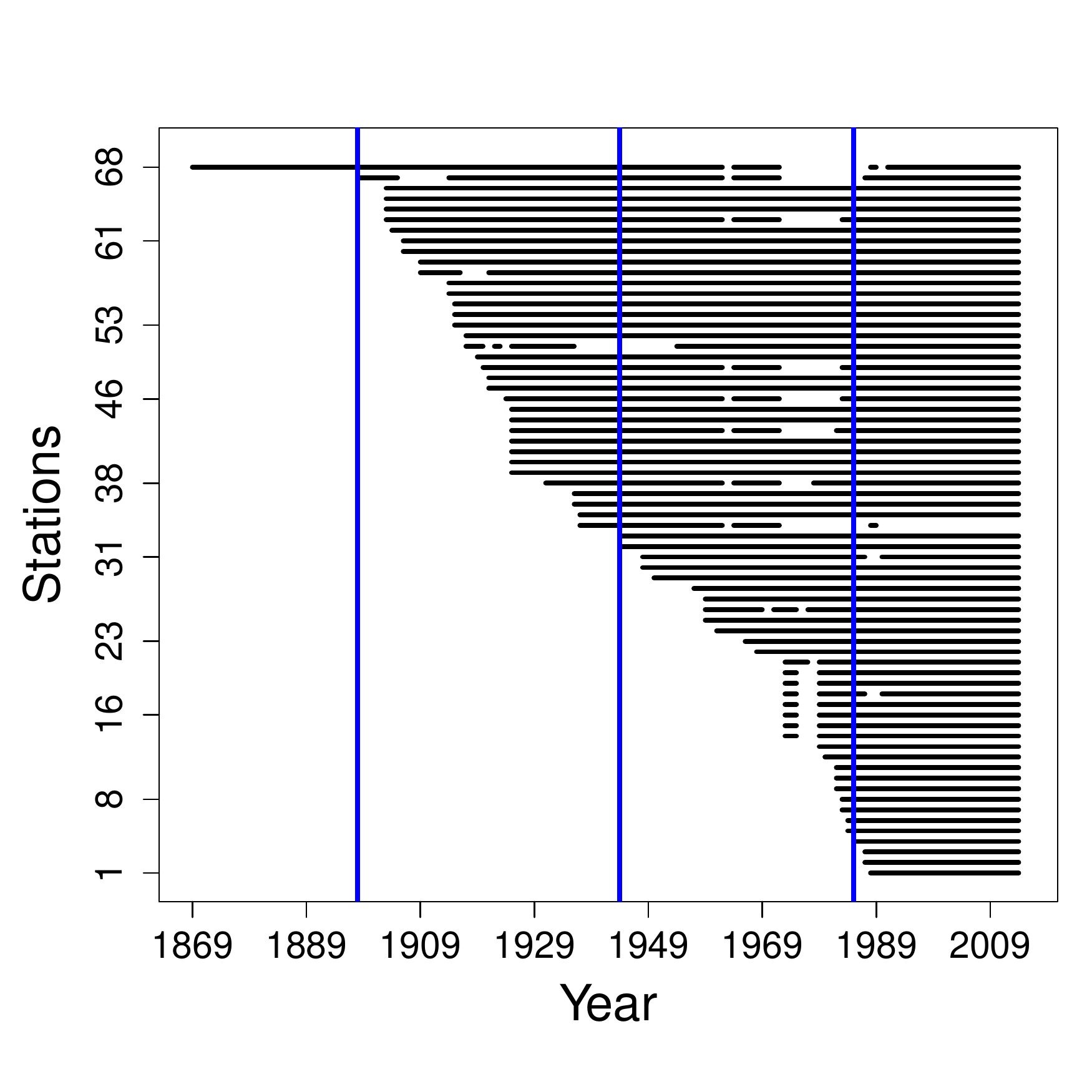}%
  \captionsetup{labelformat=empty}
}
\hspace{-1em}
\subfloat[]{%
   \includegraphics[trim = 0mm 10mm 5mm 15mm, clip,scale=.4]{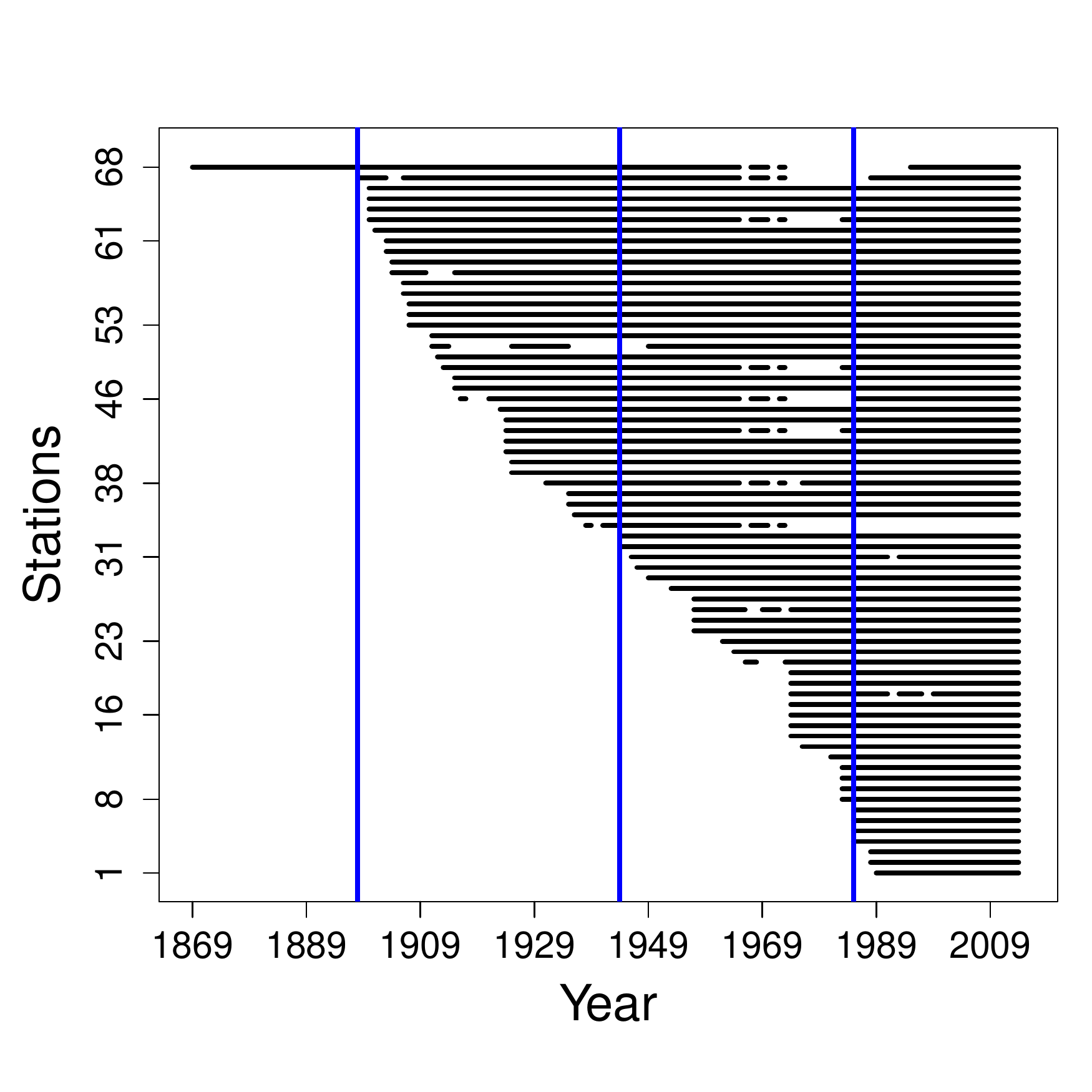}%
 \captionsetup{labelformat=empty}
}
\caption{Availability of data used for uncertainty analysis.  Top left panel: availability of original data for the 68 stations used in our analysis, with four strata shown by the blue lines.  Other panels: examples of bootstrap data sets obtained by stratified resampling of years from the original series.}
\label{Fig_DataLength}
\end{figure}

We use a nonparametric bootstrap to quantify the uncertainty of estimated quantiles obtained using our regionalization method. The upper left panel of Figure~\ref{Fig_DataLength} shows the data available for all stations.  Since the lengths of the data series differ, we partition the years from 1869 to 2015 into four strata. For each stratum we resample with equal probability and replacement from the years in the stratum. This ensures that any spatial dependence among the data at different stations is preserved. Examples of such bootstrapped data sets are shown in the other panels of Figure~\ref{Fig_DataLength}. The regional model is then fitted to the resampled data, re-estimating the optimal group size $J$ and the relative weights, and the quantiles of interest are computed.

We repeat the above bootstrapping procedure $R$ times. Then the limits of the $100(1-\alpha)\%$ bootstrap percentile confidence interval for a given station are obtained as the $\alpha/2$ and $(1-\alpha/2)$ quantiles of the $R$ bootstrap estimates. The number $R$ of bootstrap repetitions should be chosen sufficiently large \citep[Chapter~5]{Davison.Hinkley:1997}; we used $R=1000$. 

Figure~\ref{Fig_QQplots} compares results for the local and regional models for four stations  in the Rhine catchment.  The QQ-plots for the regional and local models in the right-hand panels  show quite similar fits, although the local fits use only data at single stations, whereas the regional fits use data from several stations, giving some confidence that using regional information does not greatly distort the fits at individual stations; the 95\% pointwise confidence bands are produced as described in \citet[\S4.2.4]{Davison.Hinkley:1997}. The return level plots for regional and local models in the left-hand panels of Figure~\ref{Fig_QQplots}, however, show that there can be big differences between the extrapolations from the local and regional fits.  The latter tend to give more stable extrapolations, and narrower confidence bands, especially for higher return levels.  

For some stations the local model gives unrealistically large 500-year return levels, with huge confidence intervals, whereas the corresponding
results for the regional model seem much more reasonable, owing to the reduction of
uncertainty. Figure~\ref{Fig_RLWithCI} in the Appendix shows the 50-, 100- and 200-year return levels,
with their 95\% confidence intervals, for the local and regional models for all $68$ stations
in the Rhine catchment. When the return period increases, the local model estimates and
their confidence intervals increase wildly for some stations, while the regional model gives
both more stable estimates and narrower confidence intervals, mainly due to stabilized  estimation of the shape parameter.

\subsection{Ungauged regionalization}
The main goal of the regional flood frequency analysis is to estimate high quantiles
of river discharges at ungauged target locations $t_0$.
To evaluate the performance of our proposed method for statistical regionalization, we compare the results with those obtained by clustering and canonical correlation analysis combined with classical quantile regression method; see Section~\ref{met_lit}.

In order to compute the errors between the estimates and the observed quantiles
we apply a leave-out-one procedure. For each gauged location $t_j$ from the data set, $j=1,\ldots ,68$, we assume that this location is ungauged and estimate high return
levels using the remaining stations by our regional model and the competing methods.
For our method we choose the minimal attribute weight as $\epsilon = 0.05$ and 
the minimal size of the region of influence as $J = 8$.

As is common practice, we assume that the local GEV distribution fitted to the observations at $t_j$ is the true distribution. The estimation errors are therefore compared to this baseline in terms of relative bias and relative root mean squared error. For a specific return period of $T$ years they are defined as
\begin{align*}
  \text{BIAS} &= \frac{1}{m} \sum_{j=1}^m\left( \frac{\widehat Q_{j}^T - Q_{j,L}^T}{Q_{j,L}^T} \right),\\
  \text{RMSE} &=\left\{ \frac{1}{m} \sum_{j=1}^m \left(  \frac{\widehat Q_j^T - Q_{j,L}^T}{Q_{j,L}^T}\right)^2 \right\}^{1/2},
\end{align*}
where $m=68$ is the number of gauged stations, $Q_{j,L}^T$ is the quantile obtained by the local GEV distribution at station $t_j$, and $\widehat Q_{j}^T$ is the estimate from our regional model or the competing methods.

\begin{figure}[htbp]
\centering
  \includegraphics[trim = 0mm 0mm 0mm 10mm,scale=.1, clip,width= .9\textwidth]{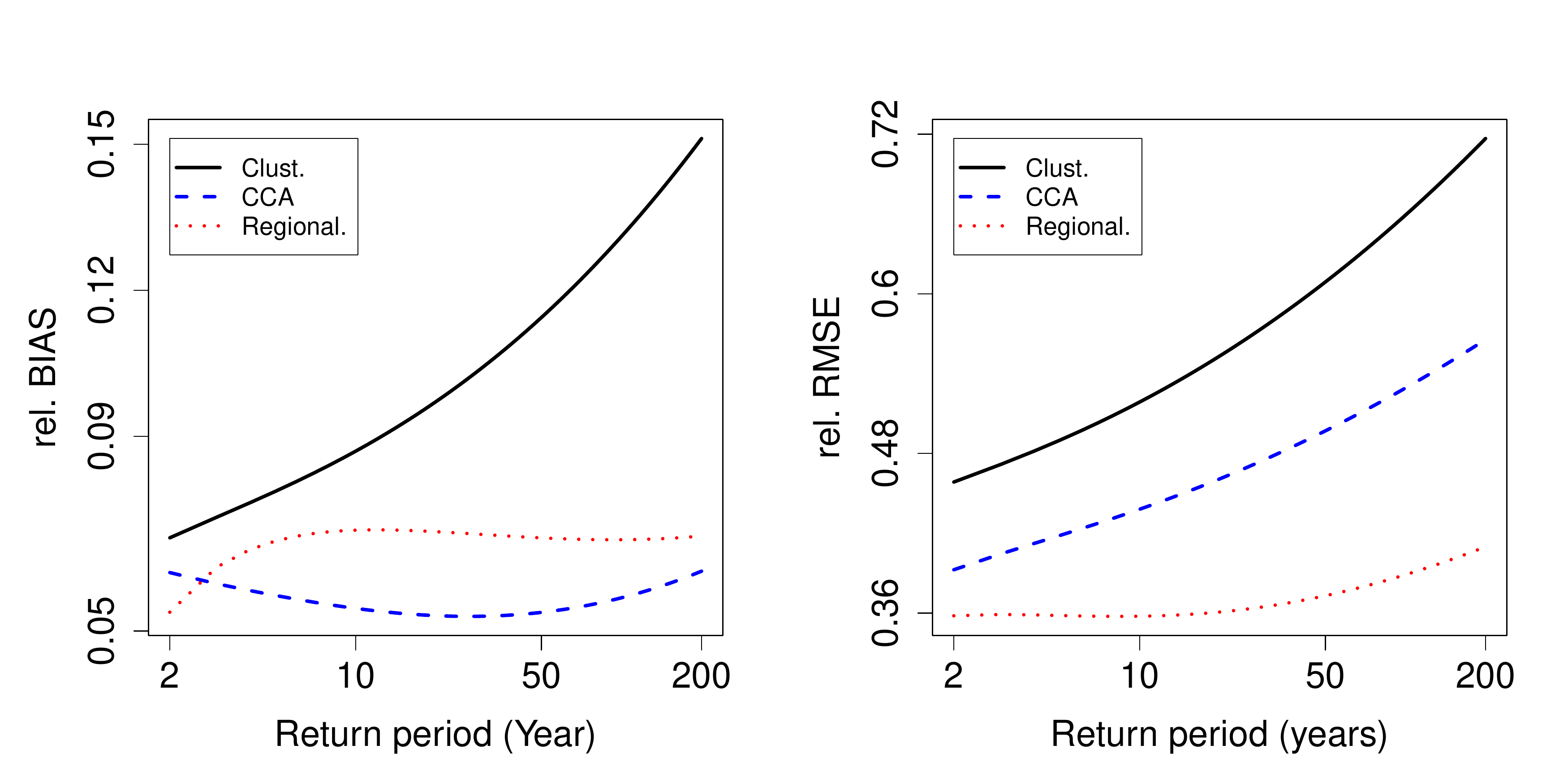}
  \caption{Relative bias and relative root mean squared error of clustering (black solid), canonical correlation analysis (blue dashed) and our regionalization method (red dotted).}
  \label{Fig_Errors}
\end{figure}

Figure~\ref{Fig_Errors} compares the relative bias and relative root mean square error  for our method and the other two methods. The relative bias of our method is smaller than that of the clustering method and almost the same as for canonical correlation analysis, whereas its relative mean squared error is much lower than those of both classical methods, especially for return periods higher than 50 years. 
The plot also shows that high quantile estimates at ungauged locations based on region of influence approaches are  much more accurate than those from the fixed-region clustering approach.

\begin{figure}[htbp]
\centering
  \includegraphics[trim = 10mm 10mm 10mm 25mm,scale=.1, clip,width= .5\textwidth]{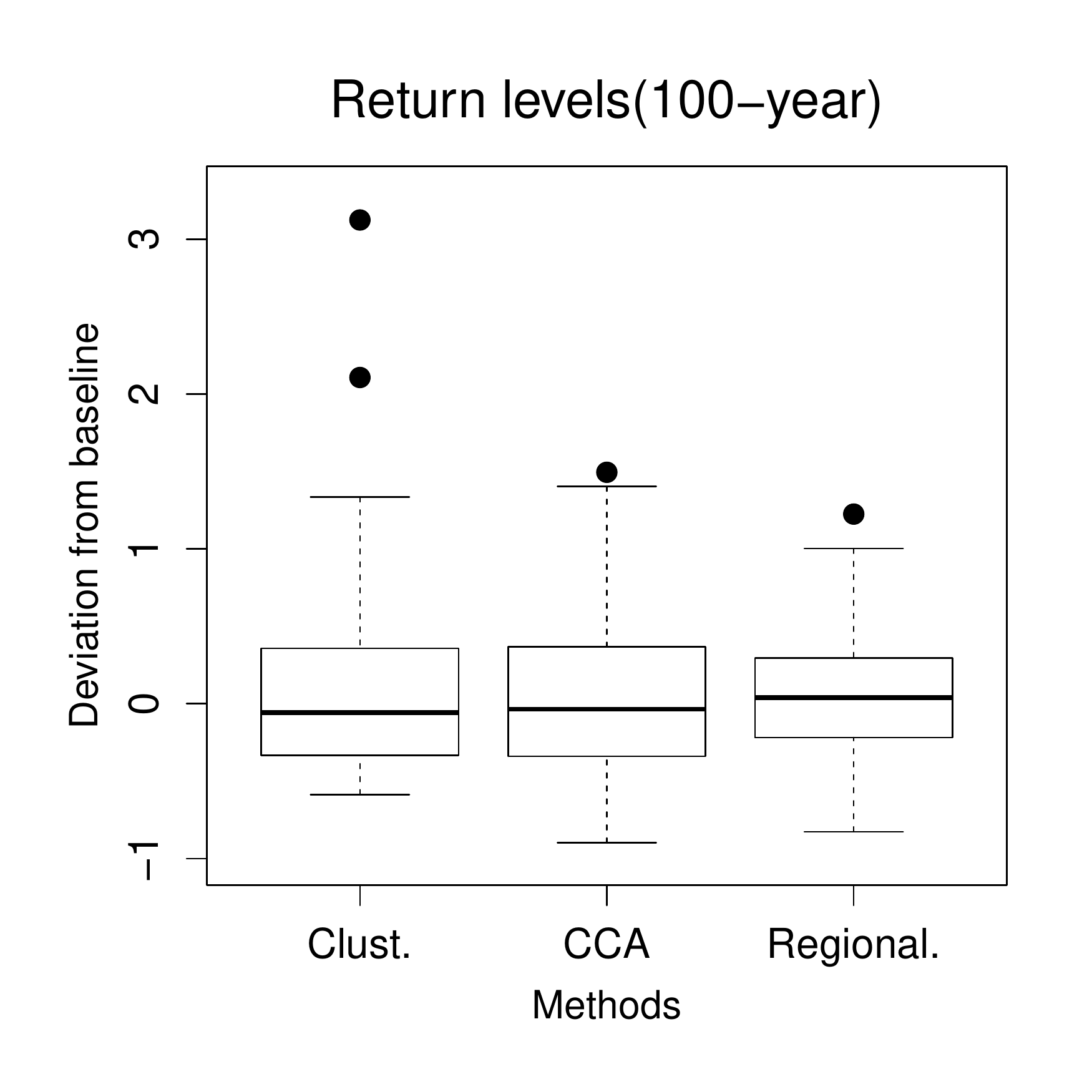}
  \caption{Deviation of 100-year return levels obtained by the three methods compared to the baseline (local model) for all 68 stations in the Rhine basin.}
  \label{Fig_BoxPlot}
\end{figure}

Figure~\ref{Fig_BoxPlot} summarizes the relative deviations of 100-year return level of the three methods from the baseline for all stations in the Rhine basin. Our statistical
regionalization has smaller variance than the competing methods and almost no outliers, so the region of influence can be chosen in a 
flexible enough way for all stations, a feature apparently missing in the fixed-region clustering.

\begin{figure}[htbp]
\centering
  \includegraphics[trim = 0mm 0mm 0mm 10mm,scale=.1, clip,width= .9\textwidth]{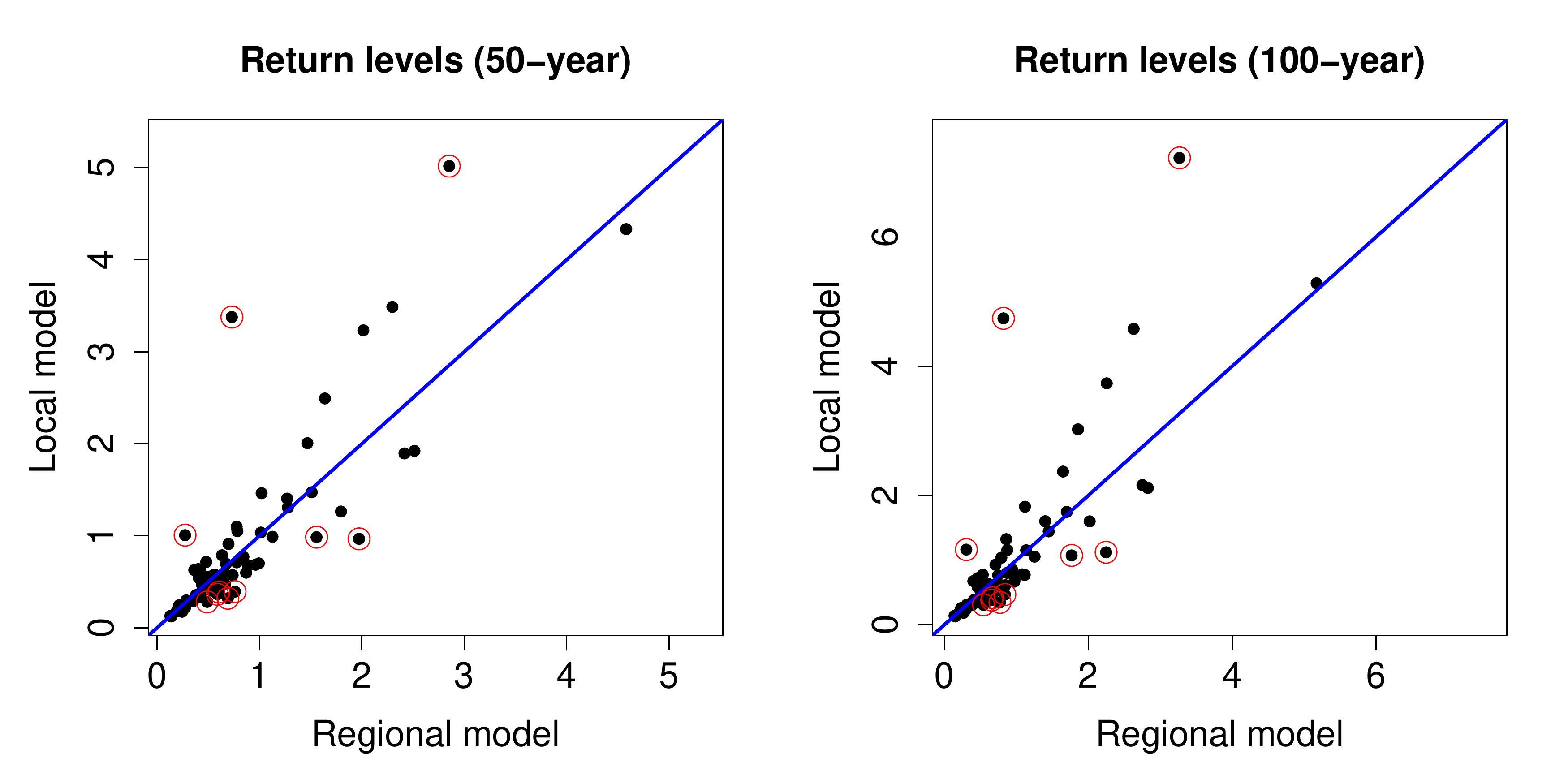}
  \caption{50- and 100-year return levels of specific discharges for the local and ungauged regional models, for the 68 stations in the Rhine river basins. Points with red circles indicate stations with big estimation errors.}
  \label{Fig_UngAndLocalQunat}
\end{figure}

Figure~\ref{Fig_UngAndLocalQunat} shows how reliably our method estimates 50-year and 100-year return levels without any
discharge data compared to the local GEV fitted to actual observations.
To be on the same scale, we plot the specific discharges, that is, the discharge at a location divided by the corresponding catchment size. 
For most stations there is very good agreement between the ungauged estimation and the
local model. The ten stations with the largest relative estimation errors, marked by red circles in Figure~\ref{Fig_UngAndLocalQunat}, mostly have 
very small catchments and no further gauging stations are located either upstream or
downstream until the next confluence point, so  there is
a lack of highly relevant information in their region of influence.

\citet{viviroli2009} studied the same region in the Rhine catchment and used 
a data base similar to ours. They implemented and compared different regionalization
methods for hydrological models and estimated high return levels
at ungauged locations. The estimation errors are not directly comparable 
with our results since the data bases do not coincide, but the orders
of magnitude seem similar.

%

\section{Conclusion}
\label{Conclusion_ROI}
The statistical regionalization approach introduced in this paper allows 
the estimation of high return levels of river discharges at both gauged and
ungauged locations. The output of the algorithm is the entire GEV distribution
of annual maxima at the point of interest. This is a major advantage over classical quantile regression techniques, since the ordering of quantiles
with different return periods is always correct. The region of influence
in our method is chosen individually for each station and in an optimal way, minimizing
the training error. This results in better performance of the model 
compared to competing methods when estimating return 
levels at ungauged locations in the Swiss Rhine basin.

For gauged locations we show that our method can improve the at-site estimation of extreme river discharges, and, in particular,
can decrease the estimation uncertainty for long return periods. This is 
achieved by using observations
in the region of influence of the target site in addition to local measurements, and thus increasing the amount of
relevant information on extreme discharges.

The small number of covariates needed in our approach and the low computation
time of the algorithm are further advantages over regionalization methods based
on hydrological models. 

\section*{Acknowledgements}
Financial support by the Swiss National Science Foundation is gratefully acknowledged.

\bibliography{References}
\small
\bibliographystyle{CUP}
\vspace{+1em}
\noindent Ecole Polytechnique F\'ed\'erale de Lausanne\\
EPFL-FSB-MATHAA-STAT\\
Station 8\\
1015 Lausanne\\
Switzerland\\
\printead{e2}\\
\phantom{E-mail:\ }\printead*{e1}\\
\phantom{E-mail:\ }\printead*{e3}
\normalsize

\newpage
\appendix
\section*{Appendix}
\vspace{+2em}
\renewcommand\thefigure{A\arabic{figure}}    
\setcounter{figure}{0} 
  
\begin{figure}[H]
\centering
  \includegraphics[trim = 10mm 10mm 5mm 5mm, clip,width= .98\textwidth]{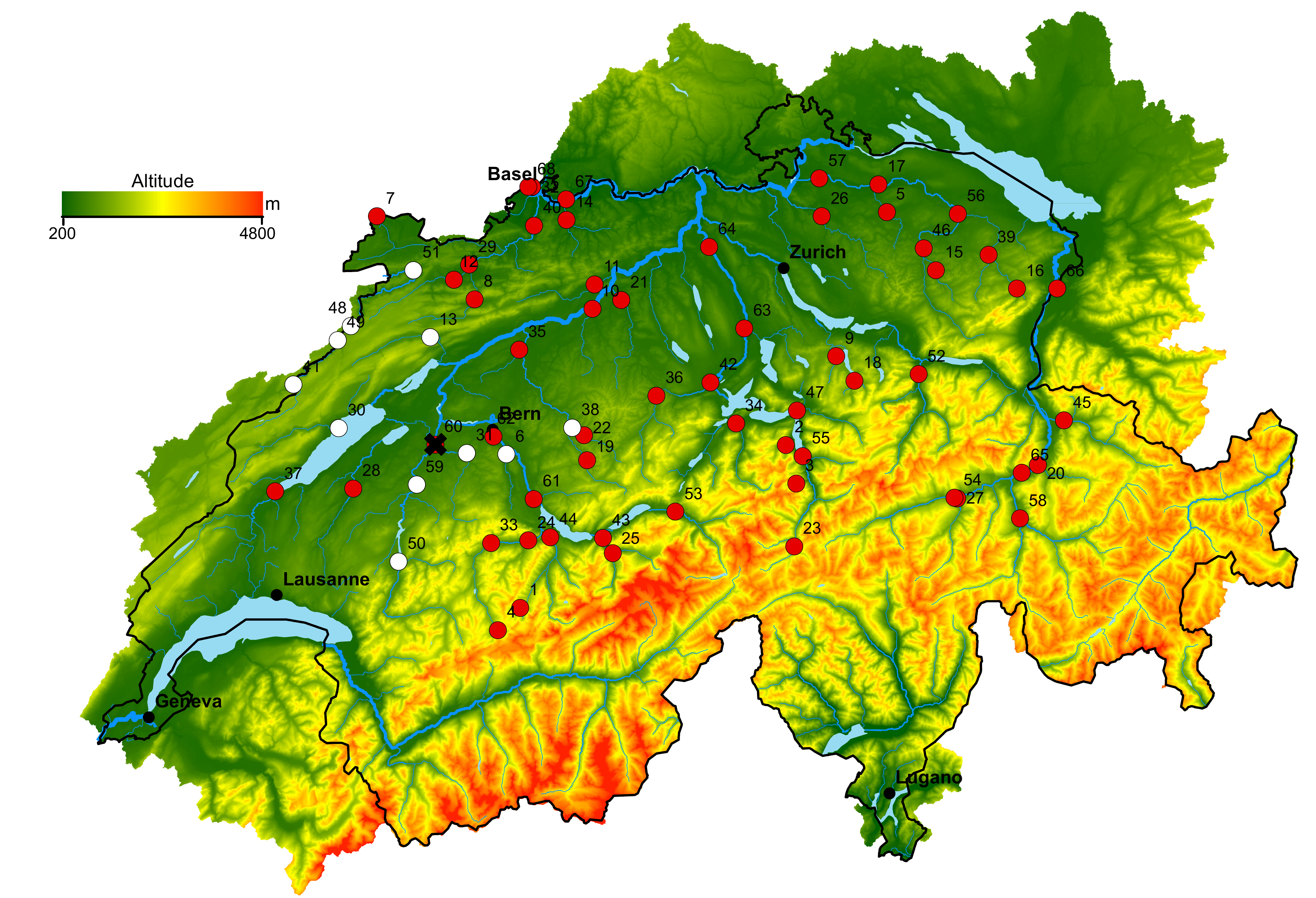}
  \caption{The white points show the stations in the optimal region of influence of station 60 (black cross).}
  \label{Fig_ROI}
\end{figure}

\begin{figure}[!t]
\captionsetup[subfigure]{labelformat=empty}
\subfloat[]{%
  \includegraphics[trim = 8mm 7mm 8mm 13mm, clip,width= .87\textwidth]{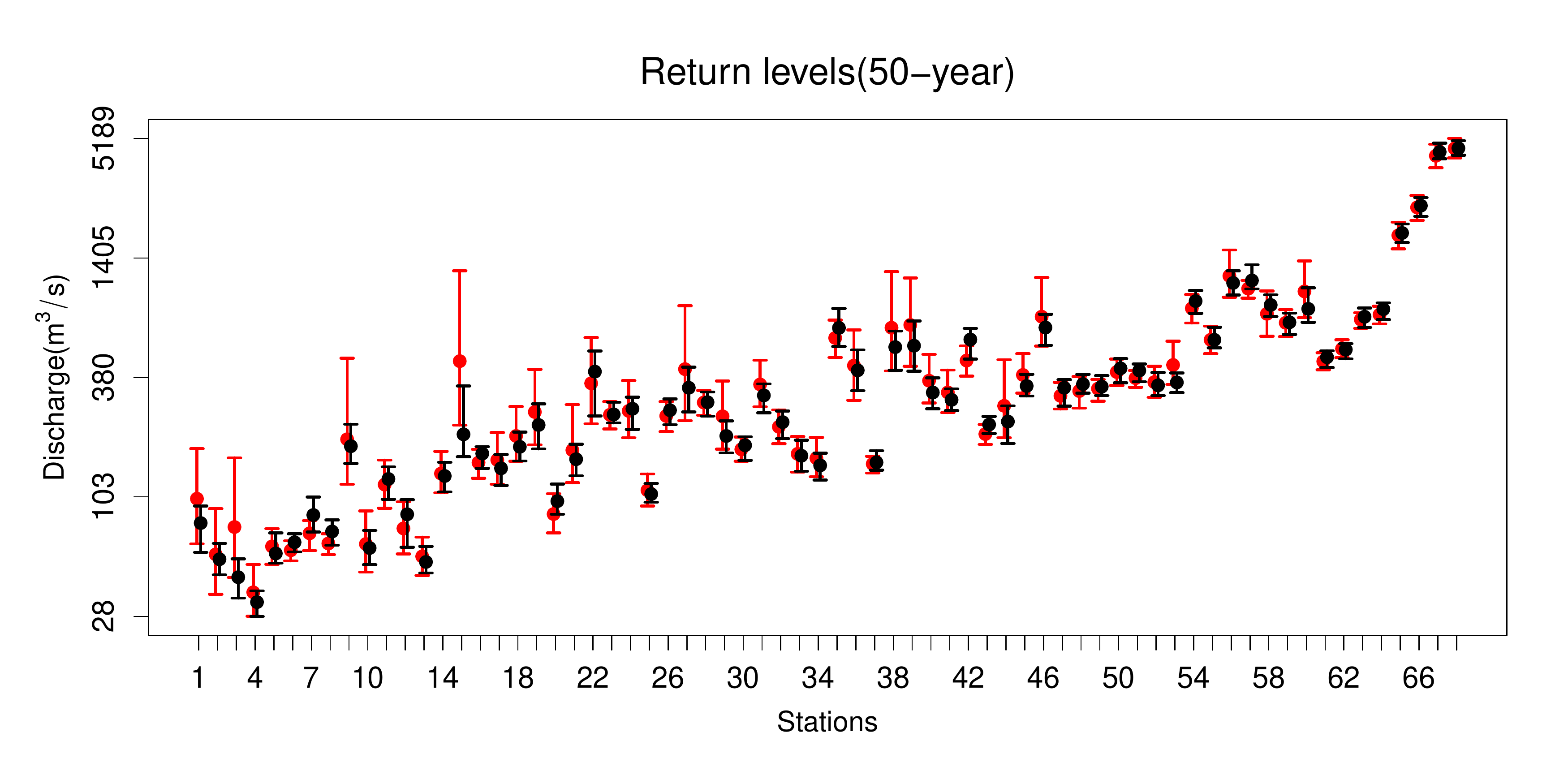}%
  \captionsetup{labelformat=empty}
}
\vspace{-1.5em}
\subfloat[]{%
  \includegraphics[trim = 8mm 7mm 8mm 10mm, clip,width= .87\textwidth]{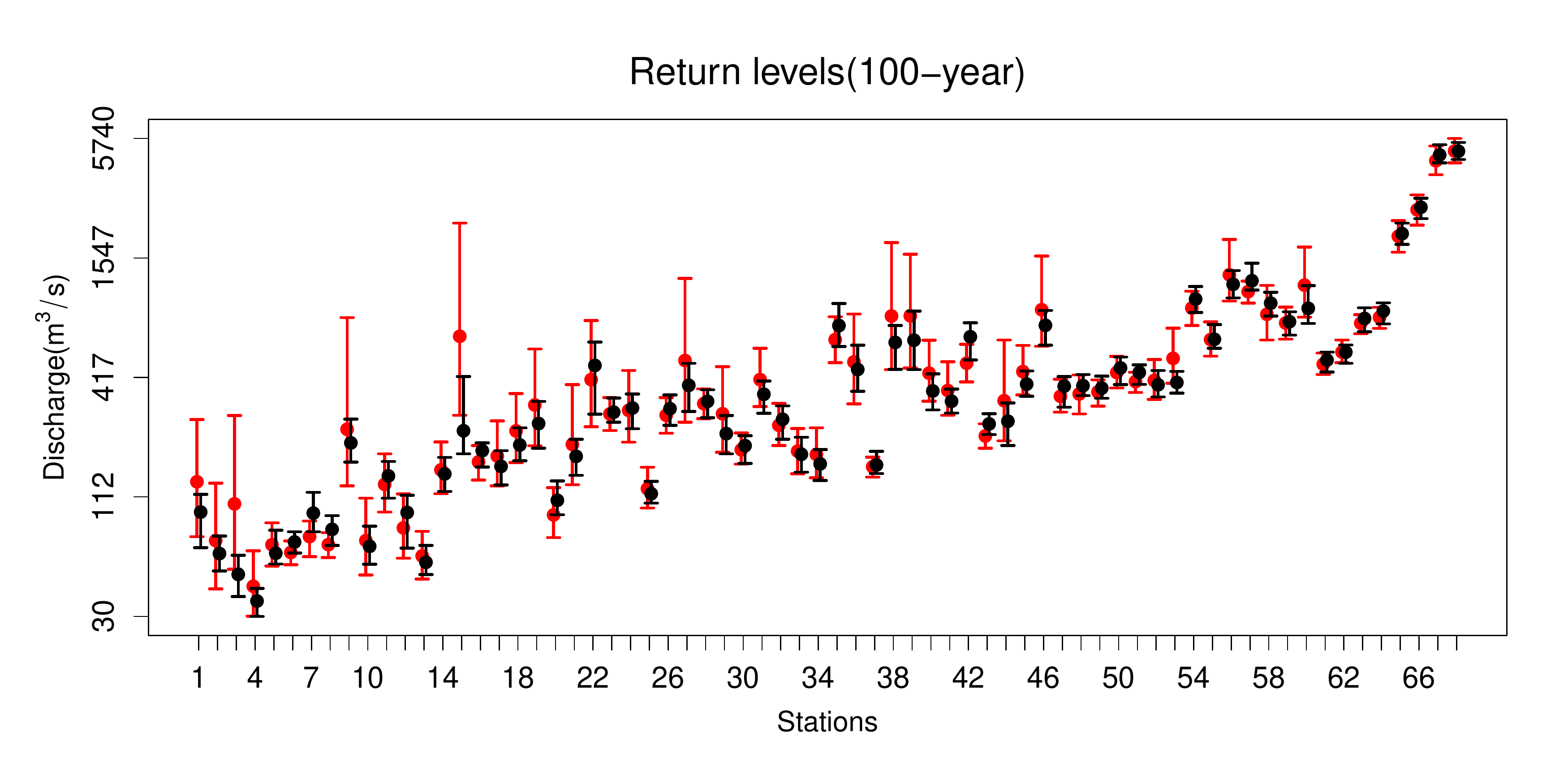}%
  \captionsetup{labelformat=empty}
}
\vspace{-1.5em}
\subfloat[]{%
  \includegraphics[trim = 8mm 7mm 8mm 10mm, clip,width= .87\textwidth]{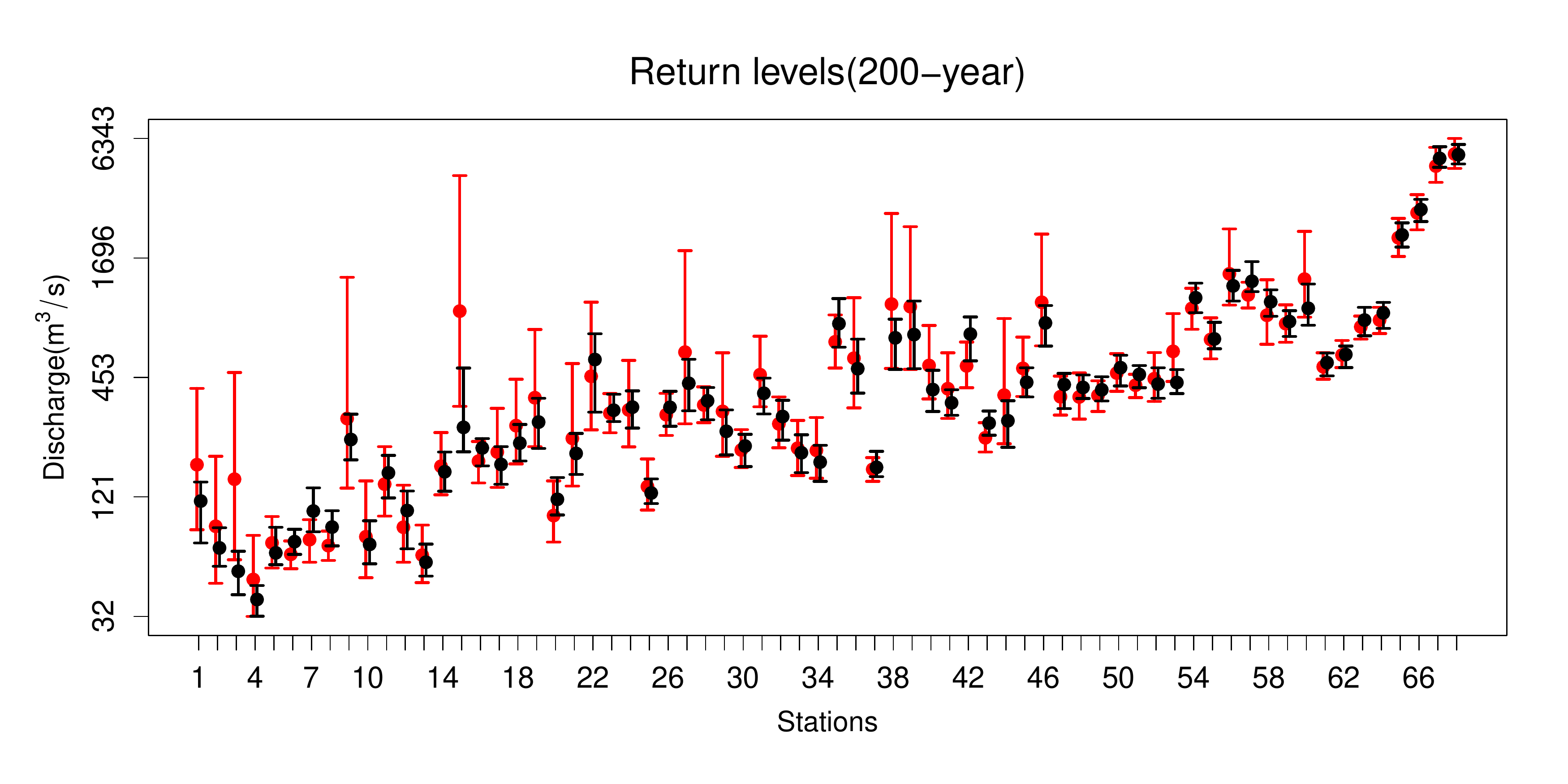}%
  \captionsetup{labelformat=empty}
}\vspace{-2em}
\caption{50-, 100- and 200-year return levels for the local(red) and regional(black) models, for the 68 stations in the Rhine river basin.  The dots show the estimates for the given return periods and the lines show the corresponding 95\% confidence intervals. The $y$-axis is logarithmic.}
\label{Fig_RLWithCI}
\end{figure} 

\end{document}